\documentclass[a4]{article}
\usepackage[T1]{fontenc}
\usepackage[latin1]{inputenc}
\usepackage{amsmath, amssymb, amsthm}
\usepackage[english]{babel}
\usepackage{indentfirst}
\usepackage{booktabs}
\usepackage{array}
\usepackage{caption,appendix}
\usepackage{geometry}
\usepackage{float}
\usepackage{cancel}
\usepackage{bbold}
\usepackage{authblk}
\numberwithin{equation}{section}
\begin{document}
\author[1,2,3,4]{Salvatore Capozziello}
\author[5]{Maurizio Capriolo} 
\author[5]{Maria Transirico}
\affil[1]{\emph{Dipartimento di Fisica "E. Pancini", Universit\`a		   di Napoli {}``Federico II'', Compl. Univ. di
		   Monte S. Angelo, Edificio G, Via Cinthia, I-80126, Napoli, Italy, }}
		  \affil[2]{\emph{INFN Sezione  di Napoli, Compl. Univ. di
		   Monte S. Angelo, Edificio G, Via Cinthia, I-80126,  Napoli, Italy,}}
 \affil[3]{\emph{Gran Sasso Science Institute, Via F. Crispi 			   	   7, I-67100, L'Aquila, Italy,}}
 \affil[4]{\emph{ Tomsk State Pedagogical University, ul. Kievskaya, 60, 634061
Tomsk, Russia, }}
\affil[5]{\emph{Dipartimento di Matematica Universit\`a di Salerno, via Giovanni Paolo II, 132, Fisciano, SA I-84084, Italy.} }
\date{\today}
\title{\textbf{The gravitational   energy-momentum  pseudo-tensor of Higher-Order Theories of Gravity}}
\maketitle
\begin{abstract}
We derive the gravitational energy momentum tensor $\tau^{\eta}_{\alpha}$ for a general Lagrangian of any order $L=L\left(g_{\mu\nu}, g_{\mu\nu,i_{1}}, g_{\mu\nu,i_{1}i_{2}},g_{\mu\nu,i_{1}i_{2}i_{3}},\cdots, g_{\mu\nu,i_{1}i_{2}i_{3}\cdots i_{n}}\right)$ and in particular for a Lagrangian such as $L_{g}=(\overline{R}+a_{0}R^{2}+\sum_{k=1}^{p} a_{k}R\Box^{k}R)\sqrt{-g}$. We prove that this tensor, in general, is not  covariant but  only affine, then it is a  pseudo-tensor. Furthermore, the   pseudo-tensor $\tau^{\eta}_{\alpha}$ is calculated in the weak field limit up to a first non-vanishing term of order $h^{2}$ where $h$ is  the metric perturbation.  The average value of the  pseudo-tensor over a suitable spacetime domain is obtained. Finally we calculate the power per unit solid angle $\Omega$ carried by a gravitational wave in a direction $\hat{x}$ for a  fixed wave number $\mathbf{k}$ under a suitable gauge. These  results are useful in view of   searching for further modes of gravitational radiation beyond the standard two modes of General Relativity and to deal with nonlocal theories of gravity where terms involving $\Box R$ are present. The general aim of the approach is to deal with theories of any order under the same standard of Landau pseudo-tensor.
\end{abstract}

\section{Introduction}
It is well know that there are many procedures for the calculation of energy momentum tensor of the gravitational field in General Relativity and many possible its definitions. This object behaves like a tensor under a linear coordinate transformation, in this case it is an affine tensor or a  pseudo-tensor, but it does not behaves like a tensor under general coordinate transformations: in this case we are dealing with a non-covariant tensor. In general,  the energy and the momentum of the gravitational field  plus matter fields are conserved in an appropriate spatially infinite region even if the gravitational energy cannot be localized. Landau and Lifshitz pointed out these features of the gravitational stress-energy tensor that, in General Relativity is a  pseudo-tensor \cite{LL}. 

However any alternative theory of gravity can present the same problem so a general method to deal with gravitational energy-momentum pseudo-tensor is needed for several practical issues ranging from the investigation of further gravitational modes in gravitational radiation, up to the identification  and classification of  nonlocal gravitational terms. In particular, such terms are now assuming a fundamental role starting from quantization of gravity up to dark energy cosmology \cite{nonlocal}.

Up to now, there is no final  quantum theory of gravity, but several  proposals indicate  the existence of  intrinsic extended structures in spacetime geometry \cite{univ4, univ5}. Such features  are always related to  effective nonlocal behaviors of spacetime and imply, in general,  higher-order terms in the effective interaction Lagrangian \cite{univ1, univ2, modesto1,modesto2}.  In  string theory, for example, 
the  measure of spacetime is impossible below the string  scale and then the theory gives rise, intrinsically,  to   effective nonlocal behaviors \cite{st1, st2}. A similar situation comes out  in loop-quantum gravity   where minimal areas have to be taken into account \cite{loop}.  In general,  any theory 
of quantum gravity  presents intrinsic extended structures comparable to the Planck length. These structures prevents to  
probe   geometry below a given scale \cite{z4,z5}.

Quantum gravitational effects produce  nonlocality and then higher order terms in the effective gravitational Lagrangian\cite{univ4, univ5,jm, jm12}.  In  cosmology, nonlocality  could be related to the  cosmic acceleration  \cite{nonl1, nonl2}. In astrophysics,    dark matter phenomena and the same  Modified Newtonian Dynamics could be explained in view of nonlocality \cite{numer, numer1}.
 
Specifically,  any nonlocal term could be represented as some $\Box R$ or $\Box^k R$ terms or similar terms in the effective gravitational Lagrangian and this feature  give rise to extensions of General Relativity. In view of these facts, it is extremely important to fix and classify   general properties of nonlocal terms, in particular the features of gravitational  stress-energy pseudo-tensor where they are involved.

The aim of this paper is to generalize the Landau Lifshitz pseudo-tensor to Extended Theories   of Gravity \cite{ET,mauro} where  Lagrangians depend on metric tensor derivatives up to $n^{th}$ order. We will show that such a  tensor, despite depending  on  metric tensor derivatives higher than second order and therefore cannot  vanish in an appropriate chart,  is an affine  non-covariant tensor, then it is a pseudo-tensor in the Landau sense. At first glance, it  might sound strange that using covariant objects, such as the tensor $g_{\mu\nu}$ or scalar densities of weight $w=-1$, like the Lagrangian $L$ and metric determinant $\sqrt{-g}$, we obtain an affine object $\tau^{\eta}_{\alpha}$ that transforms like a tensor under affinities but not under general diffeomorphisms. However there are other examples of affine object in General Relativity, e.g. affine connections $\Gamma$, that we can   call {\it pseudo-tensorial field} \cite{WP}, that not being a covariant tensor,  transforms like a tensor under affinity transformations. The reason of this behavior is  in the definition of the gravitational energy momentum tensor $\tau^{\eta}_{\alpha}$ and affine connection $\Gamma$: both of them are  functions of ordinary derivatives of the metric $g_{\mu\nu}$ and then  are non-covariant objects. Furthermore the energy of the gravitational field still is not localizable.

 However, in the weak-field limit, after  a suitable gauge choice, the gravitational energy-momentum  pseudo-tensor for a Lagrangian of $n^{th}$ order becomes a more manageable object after it has been averaged over a suitable spacetime domain. In fact, after an accurate average procedure, it would be possible to calculate the power of emitted  gravitational radiation by some astrophysical  source. The approach can be relevant in order to  investigate possible  additional polarization states of gravitational waves besides the standard two of   General Relativity \cite{greci, patrizia}. 

The paper is organized as follows. In Sec. \ref{EMTG} we obtain the gravitational energy momentum tensor for a  general Lagrangian of $n^{th}$ order in two ways: by  locally varying the gravitational Lagrangian and  by adopting the   Landau-Lifshitz procedure \cite{LL}. Hence in Sec. \ref{PNC},  we prove that this tensor is an affine tensor and not a covariant one. In Sec. \ref{EMTL} we calculate the stress energy  pseudo-tensor of gravitation field for a particular Lagrangian $L_{g}=(\overline{R}+a_{0}R^{2}+\sum_{k=1}^{p} a_{k}R\Box^{k}R)\sqrt{-g}$. Sec. \ref{EMTLO}, is devoted to the  weak-field limit of the gravitational stress-energy pseudo-tensor. We expand the gravitational energy-momentum  pseudo-tensor  in the metric perturbation $h$ up to the $h^2$ order by providing two  simple cases where the index $p$ is equal to 0 and 1.  In Sec. \ref{MVEMT},   the average value of gravitational energy-momentum  pseudo-tensor on a 4-dimensional region is calculated. We assume that the region is   large enough that any integral asymptotically  vanishes. Explicit calculations of emitted power are performed in view of applications to the gravitational waves physics.  Conclusions are drawn in Sec.\ref{conclusions}.  Finally in Appendix \ref{AppA},  we give the demonstration that the additive terms related to the symmetries of  $g_{\mu\nu}$ and its  derivatives  are on average equal to zero, i.e.   $\langle\left(A_{p}\right)_{\alpha}^{\eta}\rangle=\langle\left(B_{p}\right)_{\alpha}^{\eta}\rangle=0$, and, in Appendix \ref{AppB}, we explicitly show the six  polarization tensors related to the gravitational waves derived from higher-order theories.

\section{The gravitational energy-momentum tensor of Fourth-Order Gravity}\label{EMTG}
Let us  calculate the stress-energy tensor for a gravitational Lagrangian depending on metric tensor $g_{\mu\nu}$ and its derivatives\footnote{ The metric signature of $g_{\mu\nu}$ is $(+\ \ , -\ \ , -\ \ , -)$, Ricci tensor is defined as
$R_{\mu\nu}=R_{\ \ \mu\rho\nu}^{\rho}$ and Riemann tensor as $R_{\ \ \beta\mu\nu}^{\alpha}=\Gamma_{\beta\nu,\mu}^{\alpha}$+\ldots}
 up to fourth order  $L=L\left(g_{\mu\nu}, g_{\mu\nu,\rho}, g_{\mu\nu,\rho\lambda},g_{\mu\nu,\rho\lambda\xi}, g_{\mu\nu,\rho\lambda\xi\sigma}\right)$. This choice is due to the fact that considering all the possible curvature invariants, without $\Box$ operators into the gravitational action, the field equations results of fourth order in metric formalism (see also \cite{arturo1,stelle}).  After,  we will generalize the approach to  a gravitational Lagrangian depending on metric tensor derivatives up to $n^{th}$ order. We will derive  the stress-energy tensor both adopting the procedure by Landau \cite{LL} and varying locally the Lagrangian. 

Let us  consider the variation of the action integral with respect to metric $g_{\mu\nu}$ and coordinates $x^{\mu}$ \cite{ET, PML, FQ}:
\begin{equation}
I=\int_{\Omega}d^{4}x L \rightarrow \tilde{\delta} I=\int_{\Omega^{\prime}} d^{4}x^{\prime} L^{\prime}-\int_{\Omega}d^{4}x L=\int_{\Omega}d^{4}x \left[{\delta}L+\partial_{\mu}\left(L\delta x^{\mu}\right)\right]
\end{equation}
where $\tilde{\delta}$ means the local variation while $\delta$ means the total variation because keeps the value of coordinate $x$ fixed. For an infinitesimal transformations like:
\begin{equation}
x^{\prime\mu}=x^{\mu}+\epsilon^{\mu}\left(x\right) 
\end{equation}
we obtain the total variation of the metric tensor:
\begin{equation}
\delta g_{\mu\nu}=g^{\prime}_{\mu\nu}\left(x\right)-g_{\mu\nu}\left(x\right)=-\epsilon^{\alpha}\partial_{\alpha}g_{\mu\nu}-g_{\mu\alpha}\partial_{\nu}\epsilon^{\alpha}-g_{\nu\alpha}\partial_{\mu}\epsilon^{\alpha}
\end{equation}
 The functional variation of the metric under global transformation $\partial_{\lambda}\epsilon^{\mu}=0$ is $\delta g_{\mu\nu}=-\epsilon^{\alpha}\partial_{\alpha}g_{\mu\nu}$ and if we require  the action to be invariant under this transformation, that is $\tilde{\delta I} =0$, for a arbitrary volume of integration $\Omega$,  we get:
\begin{equation}
\begin{split}
   0=\delta L +\partial_{\mu}\left(L\delta x^{\mu}\right)=\biggl(\frac{\partial L}{\partial g_{\mu\nu}}-\partial_{\rho}\frac{\partial L}{\partial g_{\mu\nu,\rho}}+\partial_{\rho}\partial_{\lambda}\frac{\partial L}{\partial g_{\mu\nu,\rho\lambda}}-\partial_{\rho}\partial_{\lambda}\partial_{\xi}
   \frac{\partial L}{\partial g_{\mu\nu,\rho\lambda\xi}}\\+\partial_{\rho}\partial_{\lambda}\partial_{\xi}\partial_{\sigma}\frac{\partial L}{\partial g_{\mu\nu,\rho\lambda\xi\sigma}}\biggr)\delta g_{\mu\nu}+\partial_{\eta}\left(2\chi\sqrt{-g}\tau_{\alpha}^{\eta}\right)\epsilon^{\alpha}
\end{split}
\end{equation}
By imposing the constrain on metric tensor $g_{\mu\nu}$ that satisfies the Euler-Lagrange equations for a gravitational Lagrangian:
\begin{equation}
\frac{\partial L}{\partial g_{\mu\nu}}-\partial_{\rho}\frac{\partial L}{\partial g_{\mu\nu,\rho}}+\partial_{\rho}\partial_{\lambda}\frac{\partial L}{\partial g_{\mu\nu,\rho\lambda}}-\partial_{\rho}\partial_{\lambda}\partial_{\xi}\frac{\partial L}{\partial g_{\mu\nu,\rho\lambda\xi}}+\partial_{\rho}\partial_{\lambda}\partial_{\xi}\partial_{\sigma}\frac{\partial L}{\partial g_{\mu\nu,\rho\lambda\xi\sigma}}=0
\end{equation}
we get a  continuity equation:
\begin{equation}
\partial_{\eta}\left(\sqrt{-g} \tau_{\alpha}^{\eta}\right)=0
\end{equation}
for an arbitrary $\epsilon^{\alpha}$ and $\tau_{\alpha}^{\eta}$ is the stress-energy tensor for a gravitational field defined as:
\begin{equation}
\begin{split}
\tau_{\alpha}^{\eta}=\frac{1}{2\chi\sqrt{-g}}\biggl[\left(\frac{\partial L}{\partial g_{\mu\nu,\eta}}-\partial_{\lambda}\frac{\partial L}{\partial g_{\mu\nu,\eta\lambda}}+\partial_{\lambda}\partial_{\xi}\frac{\partial L}{\partial g_{\mu\nu,\eta\lambda\xi}}-\partial_{\lambda}\partial_{\xi}\partial_{\sigma}\frac{\partial L}{\partial g_{\mu\nu,\eta\lambda\xi\sigma}}\right)g_{\mu\nu,\alpha}\\
+\left(\frac{\partial L}{\partial g_{\mu\nu,\rho\eta}}-\partial_{\xi}\frac{\partial L}{\partial g_{\mu\nu,\rho\eta\xi}}+\partial_{\xi}\partial_{\sigma}\frac{\partial L}{\partial g_{\mu\nu,\rho\eta\xi\sigma}}\right)g_{\mu\nu,\alpha\rho}
+\left(\frac{\partial L}{\partial g_{\mu\nu,\rho\lambda\eta}}-\partial_{\sigma}\frac{\partial L}{\partial g_{\mu\nu,\rho\lambda\eta\sigma}}\right)g_{\mu\nu,\rho\lambda\alpha}\\
+\frac{\partial L}{\partial g_{\mu\nu,\rho\lambda\eta\sigma}}g_{\mu\nu,\rho\lambda\xi\alpha}-\delta^{\eta}_{\alpha}L\biggr]
\end{split}
\end{equation}
In a more compact form:
\begin{multline}\label{tensem4}
\tau_{\alpha}^{\eta}=\frac{1}{2\chi\sqrt{-g}}\biggl[\sum_{m=0}^{3}\left(-1\right)^{m}\left(
\frac{\partial L}{\partial g_{\mu\nu,\eta i_{0}\cdots i_{m}}}\right)_{,i_{0}\cdots i_{m}}g_{\mu\nu,\alpha}
 \\ +\sum_{j=0}^{2}\sum_{m=j+1}^{3}\left(-1\right)^{j}\left(
\frac{\partial L}{\partial g_{\mu\nu,\eta i_{0}\cdots i_{m}}}\right)_{,i_{0}\cdots i_{j}}g_{\mu\nu,i_{j+1}\cdots i_{m}\alpha}-\delta_{\alpha}^{\eta}L\biggr]
\end{multline}
where we used the following notation:
\begin{equation*}
\left(\right)_{,i_{0}}=\mathbb{1} ; \qquad \left(\right)_{,i_{0}\cdots i_{m}}=
\begin{cases}
\left(\right)_{,i_{1}}& \quad \text{if} \quad m=1\\
\left(\right)_{,i_{1}i_{2}}& \quad \text{if} \quad  m=2\\
\left(\right)_{,i_{1}i_{2}i_{3}}& \quad \text{if} \quad m=3\\
\text{and so on}&
\end{cases}
;\qquad \left(\right)_{,i_{k}\.i_{k}}=\left(\right)_{,i_{k}}
\end{equation*}

 Let us now  consider a general Lagrangian density depending up to $n^{th}$ derivative of $g_{\mu\nu}$ that is
$L=L\left(g_{\mu\nu}, g_{\mu\nu,i_{1}}, g_{\mu\nu,i_{1}i_{2}},g_{\mu\nu,i_{1}i_{2}i_{3}},\cdots, g_{\mu\nu,i_{1}i_{2}i_{3}\cdots i_{n}}\right)$. The most general total variation of $L$ and the Euler-Lagrange equations for this Lagrangian are:
\begin{equation}
  \delta L=\sum_{m=0}^{n}\frac{\partial L}{\partial g_{\mu\nu,i_{0}\cdots i_{m}}}\delta g_{\mu\nu,i_{0}\cdots i_{m}}=\sum_{m=0}^{n}\frac{\partial L}{\partial g_{\mu\nu,i_{0}\cdots i_{m}}}\partial_{i_{0}\cdots i_{m}}\delta g_{\mu\nu}
\end{equation}
\begin{equation}
  \sum_{m=0}^{n}\left(-1\right)^{m}\partial_{i_{0}\cdots i_{m}}\frac{\partial L}{\partial g_{\mu\nu,i_{0}\cdots i_{m}}}=0
\end{equation}
where $\delta g_{\mu\nu,i_{0}\cdots i_{m}}=\partial_{i_{0}\cdots i_{m}}\delta g_{\mu\nu}$ because  we are varying  keeping $x$ fixed. Proceeding in the same way as for fourth order gravity, we find a conserved quantity that satisfies a more general conservation law which allows us to define the energy momentum pseudo-tensor (which is an affine tensor as it will be proved later) for the gravitational field of $n^{th}$ order gravity
\begin{equation}\label{tensemN}
\boxed{
\begin{split}
\tau_{\alpha}^{\eta}=\frac{1}{2\chi\sqrt{-g}}\biggl[\sum_{m=0}^{n-1}\left(-1\right)^{m}\left(
\frac{\partial L}{\partial g_{\mu\nu,\eta i_{0}\cdots i_{m}}}\right)_{,i_{0}\cdots i_{m}}g_{\mu\nu,\alpha}\\+\Theta_{\left[2,+\infty\right[}\left(n\right)\sum_{j=0}^{n-2}\sum_{m=j+1}^{n-1}\left(-1\right)^{j}\left(
\frac{\partial L}{\partial g_{\mu\nu,\eta i_{0}\cdots i_{m}}}\right)_{,i_{0}\cdots i_{j}}g_{\mu\nu,i_{j+1}\cdots i_{m}\alpha}-\delta_{\alpha}^{\eta}L\biggr]
\end{split}
}
\end{equation}
where $\Theta$ is the step function:
\begin{equation}
\Theta_{\left[a,+\infty\right[}\left(n\right)=
\begin{cases}
1& \quad \text{if} \quad n\in \left[a,+\infty\right[\\
0& \quad \text{otherwise}
\end{cases}
\end{equation}
An alternative way to obtain the tensor (\ref{tensemN}) is the procedure developed by Landau \cite{LL}. We will consider, as  example,  the tensor derived from forth gravity (\ref{tensem4}):  the generalization to higher order Lagrangians is formally  identical. First of all, let us  impose the stationary condition and  vary the action with respect to the metric to find  the field equations under the hypothesis that both $\delta g_{\mu\nu}$ and the variation of derivative $\delta \partial^{n}g$ vanishing on the boundary of integration domain  to cancel the surface integrals. Hence we have:
\begin{gather}
\delta I=\delta\int_{\Omega}d^{4}x L\left(g_{\mu\nu}, g_{\mu\nu,\rho}, g_{\mu\nu,\rho\lambda},g_{\mu\nu,\rho\lambda\xi}, g_{\mu\nu,\rho\lambda\xi\sigma}\right) =0 \\
\updownarrow\nonumber\\
\frac{\partial L}{\partial g_{\mu\nu}}-\partial_{\rho}\frac{\partial L}{\partial g_{\mu\nu,\rho}}+\partial_{\rho}\partial_{\lambda}\frac{\partial L}{\partial g_{\mu\nu,\rho\lambda}}-\partial_{\rho}\partial_{\lambda}\partial_{\xi}\frac{\partial L}{\partial g_{\mu\nu,\rho\lambda\xi}}+\partial_{\rho}\partial_{\lambda}\partial_{\xi}\partial_{\sigma}\frac{\partial L}{\partial g_{\mu\nu,\rho\lambda\xi\sigma}}=0
\end{gather}
Now we calculate the  derivative and then substitute into the  field equations.  We have:
\begin{equation}
\begin{split}
\frac{\partial L}{\partial x^{\alpha}} =&\frac{\partial L}{\partial g_{\mu\nu}}\frac{\partial g_{\mu\nu}}{\partial x^{\alpha}}+\frac{\partial L}{\partial g_{\mu\nu,\rho}}\frac{\partial g_{\mu\nu,\rho}}{\partial x^{\alpha}}+\frac{\partial L}{\partial g_{\mu\nu,\rho\lambda}}\frac{\partial g_{\mu\nu,\rho\lambda}}{\partial x^{\alpha}}+\frac{\partial L}{\partial g_{\mu\nu,\rho\lambda\xi}}\frac{\partial g_{\mu\nu,\rho\lambda\xi}}{\partial x^{\alpha}}+\frac{\partial L}{\partial g_{\mu\nu,\rho\lambda\xi\sigma}}\frac{\partial g_{\mu\nu,\rho\lambda\xi\sigma}}{\partial x^{\alpha}}\\
 =&\partial_{\rho} \frac{\partial L}{\partial g_{\mu\nu,\rho}}g_{\mu\nu,\alpha}-\partial_{\rho}\partial_{\lambda}\frac{\partial L}{\partial g_{\mu\nu,\rho\lambda}}g_{\mu\nu,\alpha}+\partial_{\rho}\partial_{\lambda}\partial_{\xi}\frac{\partial L}{\partial g_{\mu\nu,\rho\lambda\xi}}g_{\mu\nu,\alpha}-\partial_{\rho}\partial_{\lambda}\partial_{\xi}\partial_{\sigma}\frac{\partial L}{\partial g_{\mu\nu,\rho\lambda\xi\sigma}}g_{\mu\nu,\alpha}\\
&+\frac{\partial L}{\partial g_{\mu\nu,\rho}} g_{\mu\nu,\rho\alpha}+\frac{\partial L}{\partial g_{\mu\nu,\rho\lambda}}g_{\mu\nu,\rho\lambda\alpha}+\frac{\partial L}{\partial g_{\mu\nu,\rho\lambda\xi}}g_{\mu\nu,\rho\lambda\xi\alpha}+\frac{\partial L}{\partial g_{\mu\nu,\rho\lambda\xi\sigma}} g_{\mu\nu,\rho\lambda\xi\sigma\alpha}\\
=&\partial_{\rho}\left(\frac{\partial L}{\partial g_{\mu\nu,\rho}}g_{\mu\nu,\alpha}\right)-\partial_{\rho}\left(\partial_{\lambda}\frac{\partial L}{\partial g_{\mu\nu,\rho\lambda}}g_{\mu\nu,\alpha}\right)+
\partial_{\lambda}\left(\frac{\partial L}{\partial g_{\mu\nu,\rho\lambda}}g_{\mu\nu,\rho\alpha}\right)+\partial_{\rho}\left(\partial_{\lambda}\partial_{\xi}\frac{\partial L}{\partial g_{\mu\nu,\rho\lambda\xi}}g_{\mu\nu,\alpha}\right)\\
&+\partial_{\lambda}\left(\frac{\partial L}{\partial g_{\mu\nu,\rho\lambda\xi}}g_{\mu\nu,\rho\xi\alpha}\right) -\partial_{\xi}\left(\partial_{\lambda}\frac{\partial L}{\partial g_{\mu\nu,\rho\lambda\xi}}g_{\mu\nu,\alpha\rho}\right)-\partial_{\rho}\left(\partial_{\lambda}\partial_{\xi}\partial_{\sigma}\frac{\partial L}{\partial g_{\mu\nu,\rho\lambda\xi\sigma}}g_{\mu\nu,\alpha}\right)\\
&+\partial_{\lambda}\left(\frac{\partial L}{\partial g_{\mu\nu,\rho\lambda\xi\sigma}}g_{\mu\nu,\rho\xi\sigma\alpha}\right)-\partial_{\xi}\left(\partial_{\lambda}\frac{\partial L}{\partial g_{\mu\nu,\rho\lambda\xi\sigma}}g_{\mu\nu,\rho\sigma\alpha}\right)+\partial_{\sigma}\left(\partial_{\xi}\partial_{\lambda}\frac{\partial L}{\partial g_{\mu\nu,\rho\lambda\xi\sigma}}g_{\mu\nu,\rho\alpha}\right)
\end{split}
\end{equation}
Grouping together terms and renaming dumb indices,  we get:
\begin{equation}
\partial_{\eta}\left(\sqrt{-g}\tau^{\eta}_{\alpha}\right)=0
\end{equation}
where $\tau^{\eta}_{\alpha}$ is the tensor defined in (\ref{tensem4}). \\
If we consider also the material Lagrangian $L_{m}=2\chi\sqrt{-g}\mathcal{L}_{m}$ whose stress-energy tensor  is defined as:
\begin{equation}
T^{\eta\alpha}=\frac{2}{\sqrt{-g}}\frac{\delta \left(\sqrt{-g}\mathcal{L}_{m}\right)}{\delta g_{\eta\alpha}}
\end{equation}
the field equations in presence of matter become  $P^{\eta\alpha}=\chi T^{\eta\alpha}$ where
\begin{equation}
P^{\eta\alpha}=-\frac{1}{\sqrt{-g}}\frac{\delta L_{g}}{\delta g_{\eta\alpha}}\,,\qquad \mbox{with the coupling} \quad \chi=\frac{8\pi G}{c^{4}}
\end{equation}
From these field equations,  we obtain:
\begin{equation}
\left(2\chi\sqrt{-g}\tau^{\eta}_{\alpha}\right)_{,\eta}=-\sqrt{-g}P^{\rho\sigma}g_{\rho\sigma,\alpha}=-\chi\sqrt{-g}T^{\rho\sigma}g_{\rho\sigma,\alpha}=2\chi\sqrt{-g} T^{\eta}_{\alpha;\eta}-\left(2\chi\sqrt{-g} T^{\eta}_{\alpha}\right)_{,\eta}
\end{equation}
\begin{equation}
\partial_{\eta}\left[\sqrt{-g}\left(\tau^{\eta}_{\alpha}+ T^{\eta}_{\alpha}\right)\right]=\sqrt{-g}T^{\eta}_{\alpha;\eta}
\end{equation}
since
\begin{equation}
\delta L +\partial_{\mu}\left(L\delta x^{\mu}\right)=-P^{\mu\nu}\sqrt{-g}\delta g_{\mu\nu}+\partial_{\eta}\left(2\chi\sqrt{-g}\tau^{\eta}_{\alpha}\right)\epsilon^{\alpha}=\left[\sqrt{-g}P^{\mu\nu}g_{\mu\nu,\alpha}+\partial_{\eta}\left(2\chi\sqrt{-g}\tau^{\eta}_{\alpha}\right)\right]\epsilon^{\alpha}=0
\end{equation}
and also because for a symmetric  tensor $T^{\eta}_{\alpha}$, one has 
\begin{equation}
\sqrt{-g}T^{\eta}_{\alpha;\eta}=\left(\sqrt{-g}T^{\eta}_{\alpha}\right)_{,\eta}-\frac{1}{2}g_{\rho\sigma,\alpha}T^{\rho\sigma}\sqrt{-g}
\end{equation}
The contracted Bianchi identities imply the conservation law of the sum of two stress-energy tensors, i.e. matter plus gravitational field, and conversely:
\begin{equation}\label{eqcontgeneralizzata}
G^{\eta\alpha}_{;\eta}=0 \leftrightarrow P^{\eta\alpha}_{;\eta}=0 \leftrightarrow T^{\eta\alpha}_{;\eta}=0 \leftrightarrow \partial_{\eta}\left[\sqrt{-g}\left(\tau^{\eta}_{\alpha}+ T^{\eta}_{\alpha}\right)\right]=0
\end{equation}
where ${\displaystyle G^{\eta\alpha}=R^{\eta\alpha}-\frac{1}{2}g^{\eta\alpha}R}$ is the Einstein tensor.
In the other worlds,   the contracted Bianchi identities  involve the conservation both of matter and gravitational field or, viceversa,  the conservation  of matter and gravitational field involves the Bianchi identities for the Einstein tensor (see also \cite{book} for a detailed discussion in alternative gravity).

From the  continuity equation (\ref{eqcontgeneralizzata}),  it is possible  to derive some conserved quantities, such as the total 4-momentum and the total angular momentum of  matter plus gravitational field,  if we assume the metric tensor derivatives up to the $n^{th}$ order vanishing on the  3-dimensional space-domain  $\Sigma$,  the  surface integral cancels out over the  boundary $\partial\Sigma$, that is   
\begin{equation}
\partial_{0}\int_{\Sigma}d^{3}x \sqrt{-g}\left(T^{\mu0}+\tau^{\mu0}\right)=-\int_{\partial\Sigma}d\sigma_{i} \sqrt{-g}\left(T^{\mu i}+\tau^{\mu i}\right)=0
\end{equation} 
where $\Sigma$ is a slice of 4-dimensional manifold of spacetime at $t$ fixed and $\partial\Sigma$ its boundary. Hence the total 4-momentum conserved is \cite{MTW}
\begin{equation}
  P^{\mu}=\int_{\Sigma}d^{3}x \sqrt{-g}\left(T^{\mu0}+\tau^{\mu0}\right)
\end{equation}
This quantity is extremely useful for practical applications in relativistic astrophysics \cite{Straumann}.

\section{Non-covariance of  gravitational energy-momentum tensor}
\label{PNC}
It is possible to prove that the tensor $\tau^{\eta}_{\alpha}$ is not generally  covariant but it is a tensor only under affine transformations \cite{WP}, that is a pseudo-tensor. Firstly we consider the particular case of the tensor (\ref{tensemN}) for $n=2$:
\begin{equation}
\tau^{\eta}_{\alpha}=\frac{1}{2\chi\sqrt{-g}}\left[\left(\frac{\partial L}{\partial g_{\mu\nu,\eta}}-\partial_{\lambda}\frac{\partial L}{\partial g_{\mu\nu,\eta\lambda}}\right)g_{\mu\nu,\alpha}+\frac{\partial L}{\partial g_{\mu\nu,\eta\xi}}g_{\mu\nu,\xi\alpha}-\delta^{\eta}_{\alpha} L\right]
\end{equation}
It is possible to  show that,  under a general diffeomorphism transformation $x^{\prime}=x^{\prime}\left(x\right)$, it is 
\begin{equation}
\tau^{\prime\eta}_{\ \alpha}\left(x^{\prime}\right) \neq \text{J}^{\eta}_{\sigma}\text{J}^{-1\tau}_{\ \ \ \alpha}\tau^{\sigma}_{\tau}\left(x\right)
\end{equation}
where the Jacobian matrix and determinant are defined as 
\begin{equation}
\text{J}^{\eta}_{\sigma}=\frac{\partial x^{\prime\eta}}{\partial x^{\sigma}}\qquad \text{J}^{-1\tau}_{\ \ \ \alpha}=\frac{\partial x^{\tau}}{\partial x^{\prime\alpha}}\qquad \text{det}\left(\text{J}^{\alpha}_{\beta}\right)=\vert J \vert =\frac{1}{\text{J}^{-1}}
\end{equation}
On the other hand, under linear or affine transformations:
\begin{equation}
x^{\prime\mu}=\Lambda^{\mu}_{\nu}x^{\nu}+a^{\mu}\qquad \text{J}^{\mu}_{\nu}=\Lambda^{\mu}_{\nu} \qquad \vert \Lambda \vert \neq 0
\end{equation}
the  tensor transforms  as
\begin{equation}
\tau^{\prime\eta}_{\ \alpha}\left(x^{\prime}\right)=\Lambda^{\eta}_{\sigma}\Lambda^{-1\tau}_{\ \ \ \alpha}\tau^{\sigma}_{\tau}\left(x\right)
\end{equation}
In general, it is possible to obtain the following identities:
\begin{equation*}
\begin{split}
\sqrt{-g^{\prime}}&=\sqrt{-g}\qquad\qquad \ \ \qquad  \text{where $g$ is a scalar density of weight $w=-2$ }\\
L^{\prime}&=\text{J}^{-1}L\qquad\qquad  \ \qquad  \text{where $L$ is a scalar density of weight $w=-1$ }\\
g^{\prime}_{\mu\nu,\alpha}\left(x^{\prime}\right)&=\text{J}^{-1 a}_{\ \ \ \mu}\text{J}^{-1 b}_{\ \ \ \nu}\text{J}^{-1 c}_{\ \ \ \alpha}g_{ab,c}\left(x\right)+\partial^{\prime}_{\alpha}\left[\text{J}^{-1 a}_{\ \ \ \mu}\text{J}^{-1 b}_{\ \ \ \nu}\right]g_{ab}\left(x\right)\\
\frac{\partial g_{\gamma\rho,\tau}}{\partial g^{\prime}_{\mu\nu,\eta}}&=\frac{1}{2}\left[\left(\delta_{a}^{\mu}\delta_{b}^{\nu}+\delta_{a}^{\nu}\delta_{b}^{\mu}\right)\delta_{c}^{\eta}\right]\text{J}_{\gamma}^{a}\text{J}_{\rho}^{b}\text{J}_{\tau}^{c}=\text{J}^{(\mu}_{\gamma}\text{J}^{\nu)}_{\rho}\text{J}^{\eta}_{\tau}\\
\frac{\partial L^{\prime}}{\partial g^{\prime}_{\mu\nu,\eta}}&=\text{J}^{-1}\text{J}^{(\mu}_{\gamma}\text{J}^{\nu)}_{\rho}\text{J}^{\eta}_{\tau}\frac{\partial L}{\partial g_{\gamma\rho,\tau}}=\text{J}^{-1}\text{J}^{\mu}_{\gamma}\text{J}^{\nu}_{\rho}\text{J}^{\eta}_{\tau}\frac{\partial L}{\partial g_{\gamma\rho,\tau}}\ \\
& \qquad\qquad\qquad\qquad\qquad\text{tensorial density (3,0) of weight $w=-1$}\\
g^{\prime}_{\mu\nu,\xi\alpha}\left(x^{\prime}\right)&=\text{J}^{-1 a}_{\ \ \ \mu}\text{J}^{-1 b}_{\ \ \ \nu}\text{J}^{-1 c}_{\ \ \ \alpha}\text{J}^{-1 d}_{\ \ \ \xi}g_{ab,cd}\left(x\right)+\partial^{\prime 2}_{\xi\alpha}\left[\text{J}^{-1 a}_{\ \ \ \mu}\text{J}^{-1 b}_{\ \ \ \nu}\right]g_{ab}\left(x\right)\\
&+\partial^{\prime}_{\alpha}\left[\text{J}^{-1 a}_{\ \ \ \mu}\text{J}^{-1 b}_{\ \ \ \nu}\right]\text{J}^{-1 d}_{\ \ \ \xi}g_{ab,d}\left(x\right)+\partial^{\prime}_{\xi}\left[\text{J}^{-1 a}_{\ \ \ \mu}\text{J}^{-1 b}_{\ \ \ \nu}\text{J}^{-1 c}_{\ \ \ \alpha}\right]g_{ab,c}\left(x\right)\\
\frac{\partial g_{\gamma\rho,\tau\epsilon}}{\partial g_{\mu\nu,\eta\xi}^{\prime}}&=\left(\delta_{a}^{(\mu}\delta_{b}^{\nu)}\delta_{c}^{(\eta}\delta_{d}^{\xi)}\right)\text{J}_{\gamma}^{a}\text{J}_{\rho}^{b}\text{J}_{\tau}^{c}\text{J}_{\epsilon}^{d}=\text{J}_{\gamma}^{(\mu}\text{J}_{\rho}^{\nu)}\text{J}_{\tau}^{(\eta}\text{J}_{\epsilon}^{\xi)}\\
\frac{\partial L^{\prime}}{\partial g^{\prime}_{\mu\nu,\eta\xi}}&=\text{J}^{-1}\text{J}^{(\mu}_{\gamma}\text{J}^{\nu)}_{\rho}\text{J}^{(\eta}_{\tau}\text{J}^{\xi)}_{\epsilon}\frac{\partial L}{\partial g_{\gamma\rho,\tau\epsilon}}=\text{J}^{-1}\text{J}^{\mu}_{\gamma}\text{J}^{\nu}_{\rho}\text{J}^{\eta}_{\tau}\text{J}^{\xi}_{\epsilon}\frac{\partial L}{\partial g_{\gamma\rho,\tau\epsilon}}\\
 & \qquad\qquad\qquad\qquad\qquad\text{tensorial density (4,0) of weight $w=-1$}\\
\partial^{\prime}_{\lambda}\frac{\partial L^{\prime}}{\partial g^{\prime}_{\mu\nu,\eta\lambda}}&=\text{J}^{-1}\text{J}^{\mu}_{\gamma}\text{J}^{\nu}_{\rho}\text{J}^{\eta}_{\tau}\text{J}^{\lambda}_{\epsilon}\text{J}^{-1 \sigma}_{\ \ \ \lambda}\partial_{\sigma}\frac{\partial L}{\partial g_{\gamma\rho,\tau\epsilon}}+\partial^{\prime}_{\lambda}\left[\text{J}^{-1}\text{J}^{\mu}_{\gamma}\text{J}^{\nu}_{\rho}\text{J}^{\eta}_{\tau}\text{J}^{\lambda}_{\epsilon}\right]\frac{\partial L}{\partial g_{\gamma\rho,\tau\epsilon}}
\end{split}
\end{equation*}
being  $A^{(\alpha\beta)}B_{\alpha\beta}=A^{\alpha\beta}B_{\alpha\beta}$,  if $B_{\alpha\beta}$ is symmetric, it is   $B_{\alpha\beta}=B_{\beta\alpha}$.
Hence we get:
\begin{equation*}
\frac{\partial L^{\prime}}{\partial g^{\prime}_{\mu\nu,\eta}}g^{\prime}_{\mu\nu,\alpha}=\text{J}^{-1}\text{J}^{\eta}_{\tau}\text{J}^{-1\pi}_{\ \ \ \alpha}\frac{\partial L}{\partial g_{\gamma\rho,\tau}}g_{\gamma\rho,\pi}\left(x\right)+\frac{\partial}{\partial x^{\prime\alpha}}\left[\text{J}^{-1a}_{\ \ \ \mu}\text{J}^{-1b}_{\ \ \ \nu}\right]g_{ab}\left(x\right)\text{J}^{-1}\text{J}^{\mu}_{\gamma}\text{J}^{\nu}_{\rho}\text{J}^{\eta}_{\tau}\frac{\partial L}{\partial g_{\gamma\rho,\tau}}\\
\end{equation*}
\begin{multline*}
\partial^{\prime}_{\lambda}\frac{\partial L^{\prime}}{\partial g^{\prime}_{\mu\nu,\eta\lambda}}g^{\prime}_{\mu\nu,\alpha}\left(x^{\prime}\right)=\text{J}^{-1}\text{J}^{\eta}_{\tau}\text{J}^{-1c}_{\ \ \ \alpha}\partial_{\sigma}\frac{\partial L}{\partial g_{ab,\tau\sigma}}g_{ab,c}+\partial^{\prime}_{\lambda}\left[\text{J}^{-1}\text{J}^{\mu}_{\gamma}\text{J}^{\nu}_{\rho}\text{J}^{\eta}_{\tau}\text{J}^{\lambda}_{\epsilon}\right]
\partial^{\prime}_{\alpha}\left[\text{J}^{-1a}_{\ \ \ \mu}\text{J}^{-1b}_{\ \ \ \nu}\right]g_{ab}\left(x\right)\frac{\partial L}{\partial g_{\gamma\rho,\tau\epsilon}}\\
+\text{J}^{-1}\text{J}^{\mu}_{\gamma}\text{J}^{\nu}_{\rho}\text{J}^{\eta}_{\tau}\partial_{\sigma}\frac{\partial L}{\partial g_{\gamma\rho,\tau\sigma}}\partial^{\prime}_{\alpha}\left[\text{J}^{-1a}_{\ \ \ \mu}\text{J}^{-1b}_{\ \ \ \nu}\right]g_{ab}+\partial^{\prime}_{\lambda}\left[\text{J}^{-1}\text{J}^{\mu}_{\gamma}\text{J}^{\nu}_{\rho}\text{J}^{\eta}_{\tau}\text{J}^{\lambda}_{\epsilon}\right]\text{J}^{-1a}_{\ \ \ \mu}\text{J}^{-1b}_{\ \ \ \nu}\text{J}^{-1c}_{\ \ \ \alpha}\frac{\partial L}{\partial g_{\gamma\rho,\tau\epsilon}}g_{ab,c}
\end{multline*}
\begin{multline*}
\frac{\partial L^{\prime}}{\partial g^{\prime}_{\mu\nu,\eta\xi}}g^{\prime}_{\mu\nu,\xi\alpha}\left(x^{\prime}\right)=\text{J}^{-1}\text{J}^{\eta}_{\tau}\text{J}^{-1\omega}_{\ \ \ \alpha}\frac{\partial L}{\partial g_{\gamma\rho,\tau\epsilon}}g_{\gamma\rho,\omega\epsilon}\left(x\right)+\text{J}^{-1}\partial^{\prime 2}_{\xi\alpha}\left[\text{J}^{-1a}_{\ \ \ \mu}\text{J}^{-1b}_{\ \ \ \nu}\right]g_{ab}\left(x\right)\text{J}^{\mu}_{\gamma}\text{J}^{\nu}_{\rho}\text{J}^{\eta}_{\tau}\text{J}^{\xi}_{\epsilon}\frac{\partial L}{\partial g_{\gamma\rho,\tau\epsilon}}\\
+\text{J}^{-1}\partial^{\prime}_{\alpha}\left[\text{J}^{-1a}_{\ \ \ \mu}\text{J}^{-1b}_{\ \ \ \nu}\right]g_{ab,d}\left(x\right)\text{J}^{\mu}_{\gamma}\text{J}^{\nu}_{\rho}\text{J}^{\eta}_{\tau}\frac{\partial L}{\partial g_{\gamma\rho,\tau d}}+\text{J}^{-1}\text{J}^{\mu}_{\gamma}\text{J}^{\nu}_{\rho}\text{J}^{\eta}_{\tau}\text{J}^{\xi}_{\epsilon}\partial^{\prime}_{\xi}\left[\text{J}^{-1a}_{\ \ \ \mu}\text{J}^{-1b}_{\ \ \ \nu}\text{J}^{-1c}_{\ \ \ \alpha}\right]g_{ab,c}\left(x\right)\frac{\partial L}{\partial g_{\gamma\rho,\tau\epsilon}}
\end{multline*}
Finally,  considering the previous relations we obtain:
\begin{equation}\label{affinitatensore}
  \tau^{\prime\eta}_{\ \alpha}\left(x^{\prime}\right)=\text{J}^{\eta}_{\sigma}\text{J}^{-1\tau}_{\ \ \ \alpha}\tau^{\sigma}_{\tau}\left(x\right)+\left\{\text{terms containing }\frac{\partial^{2} x}{\partial x^{\prime 2}},\frac{\partial^{3} x}{\partial x^{\prime 3}} \right\}
\end{equation}
This result  proves the  non-covariance of the stress-energy tensor of the gravitational field which is invariant under  affine transformations because of additional terms containing derivatives of order higher than or equal to two vanish for any  non-singular linear transformation. This result stems from the non-covariance of the derivatives of the metric tensor $g_{\mu\nu}$. Such derivatives give rise to an affine tensor.
In general,  considering 
\begin{multline*}
  g^{\prime}_{\mu\nu,i_{1}\cdots i_{m}\alpha}\left(x^{\prime}\right)=\text{J}^{-1\alpha}_{\ \ \ \mu}\text{J}^{-1\beta}_{\ \ \ \nu}\text{J}^{-1 j_{1}}_{\ \ \ i_{1}}\cdots\text{J}^{-1 j_{m}}_{\ \ \ i_{m}}\text{J}^{-1\tau}_{\ \ \ \alpha}g_{\alpha\beta,j_{1}\cdots j_{m}\tau}\left(x\right)\\
+\left\{\text{containing terms}\;\frac{\partial^{2} x}{\partial x^{\prime 2}},\cdots,\frac{\partial^{m+2}x}{\partial x^{\prime m+2}} \right\}
\end{multline*}
and
\begin{equation*}
 \frac{\partial L^{\prime}}{\partial g^{\prime}_{\mu\nu,\eta i_{0}\cdots i_{m}}}=\text{J}^{-1}\text{J}^{\mu}_{\gamma}\text{J}^{\nu}_{\rho}\text{J}^{\eta}_{\tau}\text{J}^{i_{1}}_{j_{1}}\cdots \text{J}^{i_{m}}_{j_{m}} \frac{\partial L}{\partial g_{\gamma\rho,\tau j_{1}\cdots j_{m}}}\quad\text{tensorial density (m+3,0) of weight $w=-1$}
\end{equation*}
the non-covariance of  tensor $\tau^{\eta}_{\alpha}$ comes out. On the other hand,  under affine transformations,  we have:
\begin{equation*}
\frac{\partial^{2}x}{\partial x^{\prime 2}}=\cdots=\frac{\partial^{m+2}x}{\partial x^{\prime m+2}}=0
\end{equation*}
\begin{equation*}
  \tau^{\prime\eta}_{\ \alpha}\left(x^{\prime}\right)=\Lambda^{\eta}_{\sigma}\Lambda^{-1\tau}_{\ \ \ \alpha}\tau^{\sigma}_{\tau}\left(x\right)
\end{equation*}
that is, the energy momentum tensor of gravitational field is an affine    pseudo-tensor. This result generalize to Extended Theories of Gravity the result in \cite{LL}. The affine character of the stress-energy tensor $\tau ^ {\eta} _ {\alpha}$ is related to the nonlocality of gravitational energy,  namely, gravitational energy in a finite-dimensional domain,  at a given  time,  depends on the choice of coordinate system  \cite{DI,MTW}. 
 
 It is worth noticing that the existence of  particular Lagrangians, giving rise to  energy momentum tensors  ${\tau_{\alpha}^{\eta}}$ where terms depending on derivatives in the parentheses of (\ref{affinitatensore}) as  ${\frac{\partial^{2} x}{\partial x^{\prime 2}},\cdots,\frac{\partial^{m+2}x}{\partial x^{\prime m+2}}}$   elide each other   cannot be excluded a priori. In such  cases, the energy momentum tensor ${\tau_{\alpha}^{\eta}}$ would be covariant as well as affine, becoming an effective tensor and not a pseudo-tensor.  However, due to the structure of (\ref{affinitatensore}), ${\tau_{\alpha}^{\eta}}$ is, in general,  a pseudo-tensor.

\section{The gravitational energy-momentum pseudo-tensor of  Higher than Fourth-Order Gravity}\label{EMTL}

Let us consider now higher than fourth order theories of gravity where nonlocal terms containing $\Box$ operators are present in the action. Such theories are becoming extremely interesting in supergravity and, in general, in gauge theories dealing with gravity  \cite{Modesto, Modesto1, Briscese}.  These theories are  relevant not only as effective field theories, but also as fundamental
theories. Indeed, there is at least a subclass of local higher derivative theories, the so called Lee-Wick  theories, that is unitary and super-renormalizable or finite at quantum level as demonstrated in \cite{modesto1, modesto2}.

Specifically.  
we want to calculate the energy momentum  pseudo-tensor $\tau_{\alpha}^{\eta}$ for a gravitational Lagrangian 
\begin{equation}
\label{higher}
L_{g}=(\overline{R}+a_{0}R^{2}+\sum_{k=1}^{p} a_{k}R\Box^{k}R)\sqrt{-g}
\end{equation}
where $\overline{R}$ is the linear part of the Ricci scalar $R$. Such a Lagrangian has been first considered in \cite{Quandt}. In fact, it is  possible to split the scalar curvature $R$ in linear and quadratic part: the former $\overline{R}$ depends  only on first derivative of metric tensor $g_{\mu\nu}$ while the latter depends linearly on second derivative of metric tensor as follows \cite{LL,WP,WE}
\begin{equation}
  R=R^{\star}+\overline{R}
\end{equation}
\begin{equation}
  R^{\star}=g^{\mu\nu}\left(\Gamma^{\rho}_{\mu\nu,\rho}-\Gamma^{\rho}_{\mu\rho,\nu}\right)
\end{equation}
\begin{equation}
\overline{R}=g^{\mu\nu}\left(\Gamma_{\mu\nu}^{\sigma}\Gamma_{\sigma\rho}^{\rho}-\Gamma_{\mu\sigma}^{\rho}\Gamma_{\nu\rho}^{\sigma}\right)
\end{equation}
In order to  obtain the  pseudo-tensor $\tau^{\eta}_{\alpha}$, we have to calculate the following derivatives 
\begin{equation}
  \frac{\partial L}{\partial g_{\mu\nu,\eta}}=\sqrt{-g}\left[\frac{\partial\overline{R}}{\partial g_{\mu\nu,\eta}}+\left(2a_{0}R+\sum_{k=1}^{p}a_{k}\Box^{k}R\right)\frac{\partial R}{\partial g_{\mu\nu,\eta}}+\sum_{k=1}^{p}a_{k}R\frac{\partial\Box^{k}R}{\partial g_{\mu\nu,\eta}}\right]
\end{equation}
\begin{equation}
  -\partial_{\lambda}\left(\frac{\partial L}{\partial g_{\mu\nu,\eta\lambda}}\right)=-\partial_{\lambda}\left(\sqrt{-g}\left[\left(2a_{0}R+\sum_{k=1}^{p}a_{k}\Box^{k}R\right)\frac{\partial R}{\partial g_{\mu\nu,\eta\lambda}}+\sum_{k=1}^{p}a_{k}R\frac{\partial\Box^{k}R}{\partial g_{\mu\nu,\eta\lambda}}\right]\right)
\end{equation}
\begin{equation}
\begin{split}
\sum_{m=2}^{n-1}\left(-1\right)^{m}\left(\frac{\partial L}{\partial g_{\mu\nu,\eta i_{0}\cdots i_{m}}}\right)_{,i_{0}\cdots i_{m}}=\sum_{m=2}^{n-1}\sum_{k=1}^{p}\left(-1\right)^{m}\partial_{i_{0}\cdots i_{m}}\left[\sqrt{-g}a_{k}R\frac{\partial\Box^{k}R}{\partial g_{\mu\nu,\eta i_{0}\cdots i_{m}}}\right]\\
=\sum_{k=1}^{p}\sum_{m=2}^{2p+3}\left(-1\right)^{m}\partial_{i_{0}\cdots i_{m}}\left[\sqrt{-g}a_{k}R\frac{\partial\Box^{k}R}{\partial g_{\mu\nu,\eta i_{0}\cdots i_{m}}}\right]=\sum_{k=1}^{p}\sum_{m=2}^{2k+1}\left(-1\right)^{m}\partial_{i_{0}\cdots i_{m}}\left[\sqrt{-g}a_{k}R\frac{\partial\Box^{k}R}{\partial g_{\mu\nu,\eta i_{0}\cdots i_{m}}}\right]
\end{split}
\end{equation}
where $\lambda=i_{1}$, $n=2p+4$ and
\begin{equation}
\frac{\partial\Box^{k}R}{\partial g_{\mu\nu,\eta i_{0}\cdots i_{m}}}=0 \qquad \text{if}\quad m>2k+1
\end{equation}
Hence,  one gets
\begin{equation}
\sum_{j=0}^{n-2}\sum_{m=j+1}^{n-1}\left(-1\right)^{j}\left(\frac{\partial L}{\partial g_{\mu\nu,\eta i_{0}\cdots i_{m}}}\right)_{,i_{0}\cdots i_{j}}=\sum_{h=1}^{p}\sum_{j=0}^{2p+2}\sum_{m=j+1}^{2p+3}\left(-1\right)^{j}\left(\sqrt{-g}a_{h}R\frac{\partial\Box^{h}R}{\partial g_{\mu\nu,\eta i_{0}\cdots i_{m}}}\right)_{,i_{0}\cdots i_{j}}
\end{equation}
{So, considering that $j+1\leq m \leq 2h+1$ $\rightarrow$ $j\leq 2h$, 
we finally obtain:
\begin{equation*}
\sum_{j=0}^{n-2}\sum_{m=j+1}^{n-1}\left(-1\right)^{j}\left(\frac{\partial L}{\partial g_{\mu\nu,\eta i_{0}\cdots i_{m}}}\right)_{,i_{0}\cdots i_{j}}=\sum_{h=1}^{p}\sum_{j=0}^{2h}\sum_{m=j+1}^{2h+1}\left(-1\right)^{j}\left(\sqrt{-g}a_{h}R\frac{\partial\Box^{h}R}{\partial g_{\mu\nu,\eta i_{0}\cdots i_{m}}}\right)_{,i_{0}\cdots i_{j}}
\end{equation*}
By substituting these expressions into (\ref{tensemN}),  we get the gravitational stress-energy   pseudo-tensor   for the  Lagrangian (\ref{higher})

\begin{equation}
\boxed{
\begin{split}\label{fulltensor}
\tau_{\alpha}^{\eta}=\tau_{\alpha\vert GR}^{\eta}+&\frac{1}{2\chi\sqrt{-g}}\Biggl\{\sqrt{-g}\left(2a_{0}R+\sum_{k=1}^{p}a_{k}\Box^{k}R\right)\left[\frac{\partial R}{\partial g_{\mu\nu,\eta}}g_{\mu\nu,\alpha}+\frac{\partial R}{\partial g_{\mu\nu,\eta\lambda}}g_{\mu\nu,\lambda\alpha}\right]\\ 
&-\partial_{\lambda}\left[\sqrt{-g}\left(2a_{0}R+\sum_{k=1}^{p}a_{k}\Box^{k}R\right)\frac{\partial R}{\partial g_{\mu\nu,\eta\lambda}}\right]g_{\mu\nu,\alpha}\\
&+\Theta_{\left[1,+\infty\right[}\left(p\right)\sum_{h=1}^{p}\Biggl\{\sum_{q=0}^{2h+1}\left(-1\right)^{q}\partial_{i_{0}\cdots i_{q}}\biggl[\sqrt{-g}a_{h}R\frac{\partial \Box^{h}R}{\partial g_{\mu\nu,\eta i_{0}\cdots i_{q}}}\biggl]g_{\mu\nu,\alpha}\\
&+\sum_{j=0}^{2h}\sum_{m=j+1}^{2h+1}\left(-1\right)^{j}\partial_{i_{0}\cdots i_{j}}\biggl[\sqrt{-g}a_{h}R\frac{\partial \Box^{h}R}{\partial g_{\mu\nu,\eta i_{0}\cdots i_{m}}}\biggl]g_{\mu\nu,i_{j+1}\cdots i_{m}\alpha}\Biggr\}\\
&-\delta_{\alpha}^{\eta}\left(a_{0}R^{2}+\sum_{k=1}^{p} a_{k}R\Box^{k}R\right)\sqrt{-g}\Biggr\}
\end{split}
}
\end{equation}
where we used the  notation
$
\partial_{i_{0}}=\mathbb{1}
$
and where, with $\tau_{\alpha\vert GR}^{\eta}$, we indicate the energy momentum pseudo-tensor of  General Relativity \cite{DI} as
\begin{equation}\label{tensoreGR}
\tau_{\alpha\vert GR}^{\eta}=\frac{1}{2\chi}\left(\frac{\partial \overline{R}}{\partial g_{\mu\nu,\eta}}g_{\mu\nu,\alpha}-\delta^{\eta}_{\alpha} \overline{R}\right)
\end{equation}
We replaced the term $\sqrt{-g}R$, that is a density scalar, with the effective one $\sqrt{-g}\overline{R}$ that is no longer a density scalar. This because only the effective term of the curvature $R$ contributes to the field equations and not the one that depends  on the second derivatives of the metric tensor. This consideration  makes stress-energy  pseudo-tensor easier to be integrated at fixed time $t$  over 3-dimensional spatial domains and then allows independent  calculations of  energy and momentum. These results are a straightforward generalization of those in \cite{PML}.

An important generalization of local Lagranian (\ref{higher}) can be achieved for  $p \rightarrow \infty$. In such a case, it becomes nonlocal. Studying this kind of  nonlocal  Lagrangians, exact or in weak approximation,  is a  fundamental issue for several reasons: for example,  it is worth for  regularizing  local terms which present problems with divergences. Under suitable assumptions for the coefficients   $a_{k}$ (e.g. $\sum_{k=0}^{\infty}\vert a_{k} \vert<\infty$ ) and for the domain of the linear differential operator   $ D^{p}$ 
\begin{equation}
D^{p}=\sum_{k=0}^{p}a_{k}\Box^{k}\,,
\end{equation}
which guarantee the weak or strong convergence:
\begin{equation}
\lim_{p\rightarrow\infty}\sum_{k=0}^{p}a_{k}\Box^{k}=F\left(\Box\right)\,,
\end{equation}
our local action becomes non local, i.e.
\begin{equation}
I=\int_{\Omega}d^{4}x\left[\overline{R}+RF\left(\Box\right)R\right]\sqrt{-g}\,.
\end{equation}
In this case, nonlocality makes differential operators of infinite order and then integral operators like
\begin{equation}
\Phi\left(x\right)=\int_{\Omega}d^{4}yF\left(x-y\right)R\left(x\right)=F\left(\Box\right)R\left(x\right)
\end{equation}
Let us perform now  the limit $n\rightarrow \infty$ for the energy momentum pseudo-tensor  in  (\ref{tensemN})  derived from a $n$-order Lagrangian. We want to study the behavior of the pseudo-tensor   (\ref{tensemN}) in a nonlocal  theory, that is:
\begin{equation}\label{limitinftens}
\bf{
\lim_{n\rightarrow\infty}\tau_{\alpha}^{\eta}\left(x\right)=\overline{\tau}_{\alpha}^{\eta}\left(x\right)
}
\end{equation}
Under the hypothesis that  $\tau_{\alpha}^{\eta}\left(x\right)$  transforms as an affine object, let us show that also its limit for $n\rightarrow\infty$, i.e.  $\overline{\tau}_{\alpha}^{\eta}\left(x\right)$, is an affine pseudo-tensor.   For an affine transformation 
\begin{equation}
x^{\prime\mu}=\Lambda^{\mu}_{\nu}x^{\nu}+a^{\mu}\qquad  \vert \Lambda \vert \neq 0
\end{equation}
the following affine pseudo-tensor transforms as : 
\begin{equation}\label{trasfafftens}
\tau^{\eta}_{\alpha}\left(x\right)=\Lambda^{-1\eta}_{\ \ \ \sigma}\Lambda^{\tau}_{\alpha}\tau^{\prime\sigma}_{\tau}\left(x^{\prime}\right)\,.
\end{equation}
Substituting  (\ref{trasfafftens}) in (\ref{limitinftens}), we have: 
\begin{equation}
\overline{\tau}_{\alpha}^{\eta}\left(x\right)=\lim_{n\rightarrow\infty}\Lambda^{-1\eta}_{\ \ \ \sigma}\Lambda^{\tau}_{\alpha}\tau^{\prime\sigma}_{\tau}\left(x^{\prime}\right)=\Lambda^{-1\eta}_{\ \ \ \sigma}\Lambda^{\tau}_{\alpha}\lim_{n\rightarrow\infty}\tau^{\prime\sigma}_{\tau}\left(x^{\prime}\right)=\Lambda^{-1\eta}_{\ \ \ \sigma}\Lambda^{\tau}_{\alpha}\overline{\tau}^{\prime\sigma}_{\tau}\left(x^{\prime}\right)
\end{equation}
which means that  $\overline{\tau}^{\sigma}_{\tau}\left(x\right)$  transforms as an affine object also in the limit $n\rightarrow\infty$.

\section{The weak-field limit of energy-momentum pseudo-tensor}\label{EMTLO}
The above energy-momentum pseudo-tensor (\ref{fulltensor}) derived from the Lagrangian (\ref{higher}) can be expanded in order to derive the weak-field  limit. 
For this purpose, we can write the spacetime metric as
\begin{equation}
g_{\mu\nu}=\eta_{\mu\nu}+h_{\mu\nu}\qquad\mbox{being}\quad |h_{\mu\nu}|\ll 1
\end{equation}
where $\eta_{\mu\nu}$ is the Minkowski metric and $h=\eta^{\mu\nu}h_{\mu\nu}$ is the trace of perturbation. 
We expand the energy momentum pseudo-tensor in power of $h$ to lower order considering terms up to  $h^2$.  Let us start  to expand the pseudo-tensor (\ref{tensoreGR}) in harmonic coordinates where $g^{\mu\nu}\Gamma^{\sigma}_{\mu\nu}=0$. The  linear part of the Ricci scalar $\overline{R}$ becomes:
\begin{equation}
\overline{R}=-g^{\mu\nu}\left(\Gamma^{\rho}_{\mu\sigma}\Gamma^{\sigma}_{\nu\rho}\right)
\end{equation}
that is 
\begin{equation}
\overline{R}=-\frac{1}{4}g^{\mu\nu}g^{\sigma\lambda}g^{\rho\epsilon}\left(g_{\epsilon\mu,\sigma}+g_{\epsilon\sigma,\mu}-g_{\mu\sigma,\epsilon}\right)\left(g_{\lambda\nu,\rho}+g_{\lambda\rho,\nu}-g_{\nu\rho,\lambda}\right)
\end{equation}
Approximating up to  second order  $h^{2}$, we have:
\begin{equation}
\left(\frac{\partial\overline{R}}{\partial g_{\alpha\beta,\gamma}}\right)^{\left(1\right)}\left(g_{\alpha\beta,\delta}\right)^{\left(1\right)}\stackrel{h^{2}}=\left(\frac{1}{2}h^{\alpha\beta\ \gamma}_{\ \ ,}h_{\alpha\beta,\delta}-h^{\gamma\alpha\ \beta}_{\ \ ,}h_{\alpha\beta,\delta}\right)
\end{equation}
being:
\begin{equation}
\begin{split}
\frac{\partial\overline{R}}{\partial g_{\alpha\beta,\gamma}}g_{\alpha\beta,\delta}=-\frac{1}{4}\biggl\{\left(g^{\mu\beta}g^{\sigma\alpha}g^{\epsilon\gamma}+g^{\mu\gamma}g^{\sigma\alpha}g^{\beta\epsilon}-g^{\mu\alpha}g^{\sigma\gamma}g^{\beta\epsilon}\right)\left(g_{\epsilon\mu,\sigma}+g_{\epsilon\sigma,\mu}-g_{\sigma\mu,\epsilon}\right)\\
+\left(g^{\beta\nu}g^{\gamma\lambda}g^{\rho\alpha}+g^{\gamma\nu}g^{\beta\lambda}g^{\rho\alpha}-g^{\alpha\lambda}g^{\beta\nu}g^{\rho\gamma}\right)\left(g_{\lambda\nu,\rho}+g_{\lambda\rho,\nu}-g_{\nu\rho,\lambda}\right)\biggr\}g_{\alpha\beta,\delta}
\end{split}
\end{equation}
 and also 
\begin{equation}
\overline{R}^{\left(2\right)}=-\frac{1}{4}\left(h^{\sigma\lambda}_{\ \ ,\rho}h_{\lambda\sigma,}^{\ \ \ \rho}-2h^{\sigma\lambda}_{\ \ ,\rho}h^{\rho}_{\ \lambda,\sigma}\right)
\end{equation}
Thus we substitute these terms into (\ref{tensoreGR}) and we find the explicit expression for the stress-energy  pseudo-tensor in General Relativity up to order $h^{2}$
 
 \begin{equation}
  \tau_{\alpha\vert GR}^{\eta}=\frac{1}{2\chi}\left[\frac{1}{2}h^{\mu\nu,\eta}h_{\mu\nu,\alpha}-h^{\eta\mu,\nu}h_{\mu\nu,\alpha}-\frac{1}{4}\delta_{\alpha}^{\eta}\left(h^{\sigma\lambda}_{\ \ ,\rho}h_{\lambda\sigma}^{\ \ ,\rho}-2h^{\sigma\lambda}_{\ \ ,\rho}h^{\rho}_{\ \lambda,\sigma}\right)\right]
\end{equation} 

Now in order to expand the extended gravity part  of the  pseudo-tensor (\ref{fulltensor}) to second order in $h$,  it is  sufficient to consider the following terms approximated to $h^{2}$ in harmonic gauge 
\begin{equation}\label{primoterm}
\left(2a_{0}R+\sum_{k=1}^{p}a_{k}\Box^{k}R\right)\frac{\partial R}{g_{\mu\nu,\eta\lambda}}g_{\mu\nu,\lambda\alpha}\stackrel{h^{2}}{\stackrel{\text{h.g.}}{=}}\frac{1}{4}\left(\sum_{k=0}^{p}a_{k}\Box^{k+1}h\right)h^{,\eta}_{\ \ \alpha}+\frac{1}{4}a_{0}h^{,\eta}_{\ \ \alpha}\Box h
\end{equation}
\begin{multline}\label{secondoterm}
  -\partial_{\lambda}\left[\sqrt{-g}\left(2a_{0}R+\sum_{k=1}^{p}a_{k}\Box^{k}R\right)\frac{\partial R}{\partial g_{\mu\nu,\eta\lambda}}\right]g_{\mu\nu,\alpha}\stackrel{h^{2}}{\stackrel{\text{h.g.}}{=}}a_{0}\Box h_{,\lambda}\left(h^{\lambda\eta}-\eta^{\eta\lambda}h\right)_{,\alpha}\\
  +\frac{1}{2}\sum_{k=1}^{p}a_{k}\Box^{k+1}h_{,\lambda}\left(h^{\lambda\eta}-\eta^{\lambda\eta}h\right)_{,\alpha}
\end{multline}
\begin{equation}\label{formuladermax}
  \sum_{h=1}^{p}\sum_{q=0}^{2h+1}\left(-1\right)^{q}\partial_{i_{0}\cdots i_{q}}\left[\sqrt{-g}a_{h}R\frac{\partial\Box^{h}R}{\partial g_{\mu\nu,\eta i_{0}\cdots i_{q}}}\right]g_{\mu\nu,\alpha}\stackrel{h^{2}}{\stackrel{\text{h.g.}}{=}}\frac{1}{2}\sum_{h=1}^{p}a_{h}\Box^{h+1}h_{,\lambda}\left(h^{\eta\lambda}-\eta^{\eta\lambda}h\right)_{,\alpha}+\left(A_{p}\right)_{\alpha}^{\eta}
\end{equation}
\begin{multline}\label{formuladerivatesup}
\begin{split}
  \sum_{h=1}^{p}\sum_{j=0}^{2h}\sum_{m=j+1}^{2h+1}\left(-1\right)^{j}\partial_{i_{0}\cdots i_{j}}\left[\sqrt{-g}a_{h}R\frac{\partial\Box^{h}R}{\partial g_{\mu\nu,\eta i_{0}\cdots i_{m}}}\right]g_{\mu\nu,i_{j+1}\cdots i_{m}\alpha}\stackrel{h^{2}}{\stackrel{\text{h.g.}}{=}}\frac{1}{4}\sum_{h=1}^{p}a_{h}\Box h \Box^{h} h^{,\eta}_{\ \ \alpha}\\
  +\frac{1}{2}\sum_{h=0}^{1}\sum_{j=h}^{p-1+h}\sum_{m=j+1-h}^{p}\left(-1\right)^{h}a_{m}\Box^{m-j}\left(h^{\eta\lambda}-\eta^{\eta\lambda}h\right)_{,i_{h}\alpha}\Box^{j+1-h}h_{,\lambda}^{\ \ i_{h}}+\left(B_{p}\right)_{\alpha}^{\eta}
\end{split}
\end{multline}
given that to lower order in $h$ we have:
\begin{equation}
\left(\frac{\partial R}{\partial g_{\mu\nu,\eta\lambda}}\right)^{\left(0\right)}=\frac{1}{2}\left(g^{\mu\eta}g^{\nu\lambda}+g^{\mu\lambda}g^{\nu\eta}-2g^{\mu\nu}g^{\eta\lambda}\right)^{\left(0\right)}=\frac{1}{2}\left(\eta^{\mu\eta}\eta^{\nu\lambda}+\eta^{\mu\lambda}\eta^{\nu\eta}-2\eta^{\mu\nu}\eta^{\eta\lambda}\right)
\end{equation}
\begin{equation}
  \left(\frac{\partial R}{\partial g_{\mu\nu,\eta\lambda}}\right)^{\left(0\right)}\left(g_{\mu\nu,\lambda\alpha}\right)^{\left(1\right)}=\left(h^{\lambda\eta}_{\ \ ,\lambda\alpha}-h^{,\eta}_{\ \ \alpha}\right)=\left(h^{\lambda\eta}-\eta^{\eta\lambda}h\right)_{,\lambda\alpha}\stackrel{\text{h.g.}}{=}-\frac{1}{2}h^{,\eta}_{\ \ \alpha}
\end{equation}
\begin{equation}
  \left(\frac{\partial R}{\partial g_{\mu\nu,\eta\lambda}}\right)^{\left(0\right)}\left(g_{\mu\nu,\alpha}\right)^{\left(1\right)}=\left(h^{\lambda\eta}-\eta^{\eta\lambda}h\right)_{,\alpha}
\end{equation}
\begin{multline}\label{derivatsupnonsimm}
\left(\frac{\partial\Box^{h}R}{\partial g_{\mu\nu,\eta i_{0}\cdots i_{m}}}\right)^{\left(0\right)}=\left(\frac{\partial\Box^{h}R}{\partial g_{\mu\nu,\eta i_{0}\cdots i_{q}}}\right)^{\left(0\right)}=\left(\frac{\partial\Box^{h}R}{\partial g_{\mu\nu,\eta i_{0}\cdots i_{2h+1}}}\right)^{\left(0\right)}\\=\eta^{i_{2}i_{3}}\cdots \eta^{i_{2h}i_{2h+1}}\left(\eta^{\mu i_{1}}\eta^{\nu\eta}-\eta^{\mu\nu}\eta^{\eta i_{1}}\right)+\cdots
\end{multline}\\
In Eqs. (\ref{formuladermax}), (\ref{formuladerivatesup}) and (\ref{derivatsupnonsimm}), we explicitly calculated only  the term deriving from  (\ref{derivordsupsimm}) without  considering the index permutations   ($\mu\nu$) and  $\left(\eta i_{1}\cdots i_{2h+1}\right)$. This because, taking into account terms obtained from permutations in $\left(A_{p}\right)_{\alpha}^{\eta}$ and $\left(B_{p}\right)_{\alpha}^{\eta}$, averaged on a suitable spacetime region, we obtain that  are equal to zero as we will see below. This mathematical trick is essential  to calculated the averaged gravitational energy-momentum   pseudo-tensor   and the emitted power.

Finally, by  substituting  equalities (\ref{primoterm}), (\ref{secondoterm}), (\ref{formuladermax}) and (\ref{formuladerivatesup}) into (\ref{fulltensor}),  we obtain the further extended gravity  term of   pseudo-tensor   $\tau^{\eta}_{\alpha}$ to second order, that we call $\tilde{\tau}^{\eta}_{\alpha}$, it is 

\begin{equation}
\boxed{
\begin{split}\label{total}
\tilde{\tau}_{\alpha}^{\eta}\stackrel{h^{2}}{=}\frac{1}{2\chi}\Biggl\{\frac{1}{4}\left(\sum_{k=0}^{p}a_{k}\Box^{k+1}h\right)h^{,\eta}_{\ \ \alpha}+\frac{1}{2}\sum_{t=0}^{p}a_{t}\Box^{t+1}h_{,\lambda}\left(h^{\eta\lambda}-\eta^{\eta\lambda}h\right)_{,\alpha}\\
+\frac{1}{2}\sum_{h=0}^{1}\sum_{j=h}^{p}\sum_{m=j}^{p}\left(-1\right)^{h}a_{m}\Box^{m-j}\left(h^{\eta\lambda}-\eta^{\eta\lambda}h\right)_{,\alpha i_{h}}\Box^{j+1-h}h_{,\lambda}^{\ \ i_{h}}\\
+\frac{1}{4}\sum_{l=0}^{p} a_{l}\Box^{l}\left(h^{,\eta}_{\ \ \alpha}-\Box h\delta_{\alpha}^{\eta}\right)\Box h+\Theta_{\left[1,+\infty\right[}\left(p\right)\left[\left(A_{p}\right)_{\alpha}^{\eta}+\left(B_{p}\right)_{\alpha}^{\eta}\right]\Biggr\}
\end{split}}
\end{equation}
where we have used the  conventions:
\begin{equation*}
\left(\right)_{,\alpha i_{0}}=\left(\right)_{,\alpha} \qquad h_{,\lambda}^{\ \ i_{0}}=h_{,\lambda}
\end{equation*}
Summing up we can split the total energy-momentum   pseudo-tensor   in the General Relativity part  and in the Extended Gravity part, that is
\begin{equation}
\tau_{\alpha}^{\eta}\stackrel{h^{2}}=\tau_{\alpha\vert GR}^{\eta}+\tilde{\tau}_{\alpha}^{\eta}
\end{equation}
As simple examples of  corrections related to  $\tilde{\tau}^{\eta}_{\alpha}$, let us consider the cases where the index $p$  runs up to   $0$ and up to $1$.\\ \\
For $p=0$, it is  $L_{g}=\left(\overline{R}+a_{0}R^{2}\right)\sqrt{-g}$ as in the case discussed in  \cite{PML}. We have
\begin{equation*}
\tau_{\alpha}^{\eta}\stackrel{h^{2}}=\tau_{\alpha\vert GR}^{\eta}+\tilde{\tau}_{\alpha}^{\eta}
\end{equation*}
with
\begin{equation}
\tilde{\tau}_{\alpha}^{\eta}\stackrel{h^{2}}=\frac{a_{0}}{2\chi}\left(\frac{1}{2}h^{,\eta}_{\ \ \alpha}\Box h+ h^{\eta}_{\ \lambda,\alpha}\Box h^{,\lambda}-h_{,\alpha}\Box h^{,\eta}-\frac{1}{4}\left(\Box h\right)^{2}\delta_{\alpha}^{\eta}\right)
\end{equation}
For $p=1$, that is $L_{g}=\left(\overline{R}+a_{0}R^{2}+a_{1}R\Box R\right)\sqrt{-g}$, one  gets 
\begin{equation*}
\tau_{\alpha}^{\eta}\stackrel{h^{2}}=\tau_{\alpha\vert GR}^{\eta}+\tilde{\tau}_{\alpha}^{\eta}
\end{equation*}
where
\begin{equation}
\begin{split}
\tilde{\tau}_{\alpha}^{\eta}\stackrel{h^{2}}=\frac{1}{2\chi}\Biggl\{\frac{1}{4}\left(2a_{0}\Box h+a_{1}\Box^{2}h\right)h^{,\eta}_{\ \ \alpha}+\frac{1}{2	}\left(2a_{0}\Box h_{,\lambda}+a_{1}\Box^{2}h_{,\lambda}\right)\left(h^{\eta\lambda}-\eta^{\eta\lambda}h\right)_{,\alpha}\\
+\frac{1}{2}a_{1}\Box\left(h^{\eta\lambda}-\eta^{\eta\lambda}h\right)_{,\alpha}\Box h_{,\lambda}+\frac{1}{2}a_{1}\left(h^{\eta\lambda}-\eta^{\eta\lambda}h\right)_{,\alpha}\Box^{2}h_{,\lambda}-\frac{1}{2}a_{1}\left(h^{\eta\lambda}-\eta^{\eta\lambda}h\right)_{,\sigma\alpha}\Box h_{,\lambda}^{\ \ \sigma}\\
+\frac{1}{4}a_{1}\Box h^{,\eta}_{\ \ \alpha}\Box h-\frac{1}{4}\delta_{\alpha}^{\eta}\left[a_{0}\left(\Box h\right)+a_{1}\left(\Box^{2} h\right)\right]\Box h+\left(A_{1}\right)_{\alpha}^{\eta}+\left(B_{1}\right)_{\alpha}^{\eta}\Biggr\}
\end{split}
\end{equation}
Clearly, the iteration can be performed to any $p$ introducing new contributions into dynamics.

\section{Averaging the  energy-momentum pseudo-tensor}\label{MVEMT}
Let us now  consider the solutions of  the linearized field equations in the weak-field limit in vacuum coming from the dynamics given by (\ref{higher}). From a physical point of view, such solutions are extremely important in order to calculate quantities related to the gravitational radiation.   In general, it is 

\begin{equation}
\label{wave}
h_{\mu\nu}\left(x\right)=\sum_{m=1}^{p+2}\int_{\Omega}\frac{d^{3}\mathbf{k}}{\left(2\pi\right)^{3}}\left(B_{m}\right)_{\mu\nu}\left(\mathbf{k}\right)e^{i\left(k_{m}\right)_{\alpha}x^{\alpha}}+c.c.
\end{equation}
where
\begin{equation}
\left(B_{m}\right)_{\mu\nu}\left(\mathbf{k}\right)=
\begin{cases}
C_{\mu\nu}\left(\mathbf{k}\right)& \quad \text{for}\quad m=1 \\
\frac{1}{3}\left[\frac{\eta_{\mu\nu}}{2}+\frac{\left(k_{m}\right)_{\mu}\left(k_{m}\right)_{\nu}}{k_{\left(m\right)}^{2}}\right]\text{A}_{m}\left(\mathbf{k}\right)&\quad \text{for} \quad m\geq2
\end{cases}
\end{equation}
Here "c.c." stands for the complex conjugate.
The  trace is 
\begin{equation}
\left(B_{m}\right)_{\lambda}^{\lambda}\left(\mathbf{k}\right)=
\begin{cases}
C_{\lambda}^{\lambda}\left(\mathbf{k}\right)&\quad \text{for}\quad m=1 \\
\text{A}_{m}\left(\mathbf{k}\right)&\quad \text{for} \quad m\geq2
\end{cases}
\end{equation}
and $k_{m}^{\mu}=\left(\omega_{m}, \mathbf{k}\right)$ with $k_{m}^{2}=\omega_{m}^{2}-\vert \mathbf{k} \vert ^{2}=\text{M}^{2}$ where $k_{1}^{2}=0$ and $k_{m}^{2}\neq 0$ for $m\geq 2$.
At fixed $\mathbf{k}$,  we obtain the following relations:
\begin{equation}
\begin{split}
  h^{\ \ \eta}_{,\alpha}=&2Re\left\{\sum_{j=1}^{p+2}\left(-1\right)\left(k_{j}\right)_{\alpha}\left(k_{j}\right)^{\eta}A_{j}e^{i k_{j}x}\right\}\\
  \Box^{m}h_{,\lambda}=&2 Re\left\{\left(-1\right)^{m}i\sum_{j=1}^{p+2}\left(k_{j}\right)_{\lambda}\left(k_{j}^{2}\right)^{m}A_{j}e^{ik_{j}x}\right\} \\
  \Box^{q}\left(h^{\eta\lambda}-\eta^{\eta\lambda}h\right)_{,\alpha}=&2Re\left\{\left(-1\right)^{q}i\sum_{l=1}^{p+2}\left(k_{l}\right)_{\alpha}\left(k_{l}^{2}\right)^{q}\left[\left(B_{l}\right)^{\eta\lambda}-\eta^{\eta\lambda}\left(B_{l}\right)_{\rho}^{\rho}\right]e^{ik_{l}x}\right\}\\
  \Box^{m}h_{,\lambda}^{\ \ \sigma}=&2 Re\left\{\left(-1\right)^{m+1}\sum_{j=1}^{p+2}\left(k_{j}\right)_{\lambda}\left(k_{j}\right)^{\sigma}\left(k_{j}^{2}\right)^{m}A_{j}e^{ik_{j}x}\right\} \\
  \Box^{q}\left(h^{\eta\lambda}-\eta^{\eta\lambda}h\right)_{,\sigma\alpha}=&2Re\left\{\left(-1\right)^{q+1}\sum_{l=1}^{p+2}\left(k_{l}\right)_{\sigma}\left(k_{l}\right)_{\alpha}\left(k_{l}^{2}\right)^{q}\left[\left(B_{l}\right)^{\eta\lambda}-\eta^{\eta\lambda}\left(B_{l}\right)_{\rho}^{\rho}\right]e^{ik_{l}x}\right\} \\
  \Box^{n}h=&2Re\left\{\left(-1\right)^{n}\sum_{r=2}^{p+2}\left(k_{r}^{2}\right)^{n}A_{r}e^{ik_{r}x}\right\}
\end{split}
\end{equation}
By using the following identities:
\begin{equation}
Re\{f\}Re\{g\}=\frac{1}{2}Re\{fg\}+\frac{1}{2}Re\{f\bar{g}\}
\end{equation}
\begin{equation}
  \left(k_{l}\right)_{\lambda}\left[\left(B_{l}\right)^{\eta\lambda}-\eta^{\eta\lambda}\left(B_{l}\right)_{\rho}^{\rho}\right]=-\frac{\left(k_{l}\right)^{\eta}}{2}A_{l}
\end{equation}
we can average the energy-momentum   pseudo-tensor   $\tau_{\alpha}^{\eta}$ over a region of spacetime $\Omega$ such that $\vert \Omega \vert \gg \frac{1}{\vert k\vert}$ \cite{WE}. It is worth noticing  that all integrals, including terms like $e^{i\left(k_{i}-k_{j}\right)_{\alpha}x^{\alpha}}$, approach to zero. If we assume the  harmonic gauge, after averaging and  some algebra, we  get (see Appendix \ref{AppA}):
\begin{equation}\label{medievarie}
\begin{split}
\left\langle\Box^{m}h_{,\lambda}\Box^{q}\left(h^{\eta\lambda}-\eta^{\eta\lambda}h\right)_{,\alpha}\right\rangle=&\left(-1\right)^{m+q+1}\sum_{l=2}^{p+2}\left(k_{l}\right)_{\alpha}\left(k_{l}\right)^{\eta} \left(k_{l}^{2}\right)^{\left(m+q\right)}\vert A_{l}\vert^{2} \\
\left\langle\Box^{m}h_{,\lambda}^{\ \sigma}\Box^{q}\left(h^{\eta\lambda}-\eta^{\eta\lambda}h\right)_{,\sigma\alpha}\right\rangle=&\left(-1\right)^{m+q+1}\sum_{l=2}^{p+2}\left(k_{l}\right)_{\alpha}\left(k_{l}\right)^{\eta} \left(k_{l}^{2}\right)^{\left(m+q\right)+1}\vert A_{l}\vert^{2}\\
\left\langle \Box^{q}h_{\ \alpha}^{,\eta}\Box^{m}h\right\rangle=&2\left(-1\right)^{m+q+1}\sum_{r=2}^{p+2}\left(k_{r}\right)_{\alpha}\left(k_{r}\right)^{\eta} \left(k_{r}^{2}\right)^{\left(m+q\right)}\vert A_{r}\vert^{2}\\
\left\langle\Box^{m}h\Box h\right\rangle=&2\left(-1\right)^{m+1}\sum_{j=2}^{p+2}\left(k_{j}^{2}\right)^{m+1}\vert A_{j}\vert^{2}\\
\langle\left(A_{p}\right)_{\alpha}^{\eta}\rangle=&\langle\left(B_{p}\right)_{\alpha}^{\eta}\rangle=0
\end{split}
\end{equation}
A base for the linearized solutions $h_{\mu\nu}$ is given in Appendix \ref{AppB}.
Using these equalities,  we can calculate the average value of the energy momentum   pseudo-tensor  :

\begin{equation}\label{MEMT}
\boxed{
\begin{split}
\left\langle\tau_{\alpha}^{\eta}\right\rangle=\frac{1}{2\chi}\left[\left(k_{1}\right)^{\eta}\left(k_{1}\right)_{\alpha}\left(C^{\mu\nu}C_{\mu\nu}^{*}-\frac{1}{2}\vert C_{\lambda}^{\lambda}\vert^{2}\right)\right]\\
+\frac{1}{2\chi}\left[\left(-\frac{1}{6}\right)\sum_{j=2}^{p+2}\left(\left(k_{j}\right)^{\eta}\left(k_{j}\right)_{\alpha}-\frac{1}{2}k_{j}^{2}\delta_{\alpha}^{\eta}\right)\vert A_{j}\vert^{2}\right]\\
+\frac{1}{2\chi}\Biggl\{\Biggl[\sum_{l=0}^{p}\left(l+2\right)\left(-1\right)^{l}a_{l}\sum_{j=2}^{p+2}\left(k_{j}\right)^{\eta}\left(k_{j}\right)_{\alpha}\left(k_{j}^{2}\right)^{l+1}\vert A_{j}\vert^{2}\Biggr]\\
-\frac{1}{2}\sum_{l=0}^{p}\left(-1\right)^{l}a_{l}\sum_{j=2}^{p+2}\left(k_{j}^{2}\right)^{l+2}\vert A_{j}\vert^{2}\delta_{\alpha}^{\eta}\Biggr\}
\end{split}
}
\end{equation}
where, as above,  ${\displaystyle \chi=\frac{8\pi G}{c^{4}}}$.
Assuming the  TT gauge for the first mode  concerning $k_{1}$ and only harmonic gauge for residual  modes $k_{m}$, it is:
\begin{equation}
\begin{cases}
  \left(k_{1}\right)_{\mu}C^{\mu\nu}=0 \quad \land \quad C_{\lambda}^{\lambda}=0&\quad \text{if}\quad m=1\\
  \left(k_{m}\right)_{\mu}\left(B_{m}\right)^{\mu\nu}=\frac{1}{2}\left(B_{m}\right)_{\lambda}^{\lambda}k^{\nu}& \quad  \text{if} \quad m\geq2
\end{cases}
\end{equation}
With these considerations in mind,  we can consider a gravitational wave traveling in the $+z$-direction at $\mathbf{k}$ fixed, with 4-wave vector given by $k^{\mu}=\left(\omega,0,0,k_{z}\right)$ where $\omega_{1}^{2}=k_{z}^{2}$ if  $k_{1}^{2}=0$ and $ k_{m}^{2}=m^{2}=\omega_{m}^{2}-k_{z}^{2}$ otherwise with $k_{z}>0$. Therefore we get the averaged tensorial component 
\begin{multline}
  \left\langle\tau_{0}^{3}\right\rangle=\frac{c^{4}}{8\pi G}\omega_{1}^{2}\left(C_{11}^{2}+C_{12}^{2}\right)+\frac{c^{4}}{16\pi G}\Biggl[\left(-\frac{1}{6}\right)\sum_{j=2}^{p+2}\omega_{j}k_{z}\vert A_{j}\vert^{2}\\
  +\sum_{l=0}^{p}\left(l+2\right)\left(-1\right)^{l}a_{l}\sum_{j=2}^{p+2}\omega_{j}k_{z}m_{j}^{2\left(l+1\right)}\vert A_{j}\vert^{2}\Biggr]
\end{multline}

As an application of these results, we can compute  the emitted power per unit solid angle $\Omega$, radiated in a direction $\hat{x}$ at a fixed $\mathbf{k}$.  Under a suitable gauge, it  is:
\begin{equation}
  \frac{dP}{d\Omega}=r^2\hat{x}^{i}\left\langle\tau_{0}^{i}\right\rangle
\end{equation}
Specific cases can be considered according to the index $p$ of the   pseudo-tensor   (\ref{total}). We have the cases
\\
p=0
\begin{gather*}
  \left\langle\tau_{0}^{3}\right\rangle=\frac{c^4\omega_{1}^{2}}{8\pi G}\left[C_{11}^{2}+C_{12}^{2}\right]+\frac{c^{4}}{16\pi G}\biggl\{\left(-\frac{1}{6}\right)\omega_{2}\vert A_{2}\vert^{2}k_{z}+2a_{0}\omega_{2}m_{2}^{2}\vert A_{2}\vert^{2}k_{z}\biggr\}
\end{gather*}
p=1
\begin{multline}
  \left\langle\tau_{0}^{3}\right\rangle=\frac{c^4\omega_{1}^{2}}{8\pi G}\left[C_{11}^{2}+C_{12}^{2}\right]+\frac{c^{4}}{16\pi G}\biggl\{\left(-\frac{1}{6}\right)\left(\omega_{2}\vert A_{2}\vert^{2}+\omega_{3}\vert A_{3}\vert^{3}\right)k_{z}\\
  +2a_{0}\left[\left(\omega_{2}m_{2}^{2}\vert A_{2}\vert^{2}+
  \omega_{3}m_{3}^{2}\vert A_{3}\vert^{2}\vert^{2}\right)k_{z}\right] -3a_{1}\left[\left(\omega_{2}m_{2}^{4}\vert A_{2}\vert^{2}+\omega_{3}m_{3}^{4}\vert A_{3}\vert^{2}\right)k_{z}\right]\biggr\}
\end{multline}
p=2
\begin{multline}
  \left\langle\tau_{0}^{3}\right\rangle=\frac{c^4\omega_{1}^{2}}{8\pi G}\left[C_{11}^{2}+C_{12}^{2}\right]+\frac{c^{4}}{16\pi G}\biggl\{\left(-\frac{1}{6}\right)\left(\omega_{2}\vert A_{2}\vert^{2}+\omega_{3}\vert A_{3}\vert^{3}+\omega_{4}\vert A_{4}\vert^{2}\right)k_{z}\\+2a_{0}\left[\left(\omega_{2}m_{2}^{2}\vert A_{2}\vert^{2}+\omega_{3}m_{3}^{2}\vert A_{3}\vert^{2}+\omega_{4}m_{4}^{2}\vert A_{4}\vert^{2}\right)k_{z}\right]\\
   -3a_{1}\left[\left(\omega_{2}m_{2}^{4}\vert A_{2}\vert^{2}+\omega_{3}m_{3}^{4}\vert A_{3}\vert^{2}+\omega_{4}m_{4}^{4}\vert A_{4}\vert^{2}\right)k_{z}\right]\\
  +4a_{2}\left[\left(\omega_{2}m_{2}^{6}\vert A_{2}\vert^{2}+\omega_{3}m_{3}^{6}\vert A_{3}\vert^{2}+\omega_{4}m_{4}^{6}\vert A_{4}\vert^{2}\right)\right]\biggr\}
\end{multline}
where we have explicitly indicated the coupling $\chi$. By a rapid inspection of the above formulas, it is clear that the first term is the General Relativity contribution while the corrections strictly depends on $p$. In any  context where corrections to General Relativity can be investigated, this approach could constitute a paradigm to search for higher order effects.
\section{Conclusions}
\label{conclusions}
Corrections to the standard Hilbert-Einstein Lagrangian and, in particular nonlocal terms,  are gaining more and more interest in view of addressing gravitational phenomena at ultraviolet and infrared scales. However most of the main features of General Relativity should be retained to get self-consistent theories. In particular, the properties of the gravitational energy momentum tensor  need a detailed consideration in view of both  foundation and applications of any gravitational theory.

Here, we derived  the gravitational energy momentum tensor $\tau^{\eta}_{\alpha}$ for a  general Lagrangian of the form  
$L=L\left(g_{\mu\nu}, g_{\mu\nu,i_{1}}, g_{\mu\nu,i_{1}i_{2}},g_{\mu\nu,i_{1}i_{2}i_{3}},\cdots, g_{\mu\nu,i_{1}i_{2}i_{3}\cdots i_{n}}\right)$ of any order in metric derivatives, showing that it is, in general, a   pseudo-tensor  .  In particular,  we considered  a Lagrangians like $L_{g}=(\overline{R}+a_{0}R^{2}+\sum_{k=1}^{p} a_{k}R\Box^{k}R)\sqrt{-g}$, in the weak field   limit up to the order  $h^2$ in metric perturbations.  Under suitable gauge conditions,  we averaged it over a suitable  four-dimensional domain. The averaged tensor (\ref{MEMT}) depends on the free parameters  $a_{m}$ and $p$, on the  amplitudes $\text{A}_{j}\left(\mathbf{k}\right)$, $C_{11}\left(\mathbf{k}\right)$ and $C_{22}\left(\mathbf{k}\right)$ other then on  the wave numbers $k_{m}^{2}=\text{M}^{2}$ that, in turn, are  linked to $a_{m}$.  As a general result, the gravitational wave \eqref{wave} associated with higher order Lagrangians can be expressed, under a suitable gauge choice  for a wave propagating along the  $+z$-direction, in terms of six polarization tensors  (see Appendix \ref{AppB}) as 
\begin{multline}
\label{GW1}
h_{\mu\nu}\left(t;z\right)=\text{A}^{\left(+\right)}\left(t-z\right)\epsilon_{\mu\nu}^{\left(+\right)}
+\text{A}^{\left(\times\right)}\left(t-z\right)\epsilon_{\mu\nu}^{\left(\times\right)}+\text{A}^{\left(TT\right)}\left(t-v_{G_{m}}z\right)\epsilon_{\mu\nu}^{\left(TT\right)}\\
+\text{A}^{\left(TS\right)}\left(t-v_{G_{m}}z\right)\epsilon_{\mu\nu}^{\left(TS\right)}
+\text{A}^{\left(1\right)}\left(t-v_{G_{m}}z\right)\epsilon_{\mu\nu}^{\left(1\right)}+\text{A}^{\left(L\right)}\left(t-v_{G_{m}}z\right)\epsilon_{\mu\nu}^{\left(L\right)}
\end{multline}
where $v_{G_{m}}$ is the group velocity of the $m_{th}$ massive mode (see also \cite{greci, arturo1}).
By using the total conservation of gravitational and matter energy momentum tensor (\ref{eqcontgeneralizzata}), the amplitudes $\text{A}_{j}\left(\mathbf{k}\right)$, $C_{11}\left(\mathbf{k}\right)$ and $C_{22}\left(\mathbf{k}\right)$ and then the  averaged gravitational tensor (\ref{MEMT}) can be expressed in terms of components of matter energy momentum tensor  $T^{\mu\nu}$ (as $T^{00}$) and then to  the source  generating  gravitational waves. In   principle, it would be possible to find the gravitational power emitted by a local astrophysical source in slow-motion approximation in terms of $T^{00}$, and then observationally fix the range of parameters  $a_{m}$ and $p$ of compatible models. The decreasing of timing of the Hulse-Taylor binary pulsar \cite{HT} is an example in this sense.
See also \cite{felicia,ivan} for  $f(R)$ gravity. This procedure could lead to fix the derivative degree $2p+4$ of the theory of gravity \cite{Quandt,staro} and could constitute a test bed for any Extended Theory of Gravity\footnote{It is worth noticing that the index $p=0$ gives, in general,  a fourth-order theory like $f(R)$. The only second-order theory is the General Relativity which is the particular case $f(R)=R$.}. In general, this procedure allows to  investigate additional polarization states of gravitational waves apart the  standard   polarizations $\epsilon_{\mu\nu}^{\left(+\right)}$ and $\epsilon_{\mu\nu}^{\left(\times\right)}$ of General Relativity. Furthermore,  the {\bf pseudo-tensor } $\left\langle\tau_{\alpha}^{\eta}\right\rangle$ depends on  $k_{m}^{2}$ which are related to massive modes: this fact   could enable to fix lower and upper limits for  massive modes. Finally, considering the average value $\left\langle\tau_{\alpha}^{\eta}\right\rangle$, it is possible to  calculate the power carried by  gravitational waves emitted by a binary black hole merger and to compare the results  with measurements already performed by LIGO-VIRGO collaboration \cite{LIGO}. Specifically, this procedure allows  to fix the degree $2p+4$ of the theory and may provide the "signature" for  further polarizations. 
In a forthcoming paper, astrophysical applications of the above results will be considered.

 As a final remark, it is important to emphasize again that the gravitational energy momentum tensor $\tau^{\eta}_{\alpha}$  is a   pseudo-tensor. This object, as the affine connections $\Gamma^{\mu}_{\nu\lambda}$, behaves like tensor components only under linear coordinate transformations but not under the general group of coordinate transformations. Even if the class of diffeomorphisms is more general  than the one of affine transformations, that is affinities are a subset of diffeomorphisms, the set of affine objects in more general than the one of covariant objects:  this  means that an object that behaves like a tensor under general diffeomorphism  behaves as a tensor also under affine transformations while the reverse is not always true. In other words, the   pseudo-tensor   $\tau^{\eta}_{\alpha}$ loses the general covariance although it is defined starting from  covariant objects. This fact is related to the presence of  partial  derivatives of the metric tensor that are not tensors. Finally, the affine properties of  the gravitational energy momentum   pseudo-tensor   could be relevant to discriminate between the teleparallel and  metric formulations of theories of gravity. For a detailed discussion of this point see Ref.\cite{manos}.
 
 \section*{Acknowledgements}
SC acknowledges INFN Sez. di Napoli (Iniziative Specifiche QGSKY and TEONGRAV) and  the COST Action CA15117 (CANTATA). 
 
\appendix
\section{ The average of  $\langle\left(A_{p}\right)_{\alpha}^{\eta}\rangle$  and $\langle\left(B_{p}\right)_{\alpha}^{\eta}\rangle$ terms} \label{AppA}
Let us now demonstrate the last two relations in  (\ref{medievarie}), that is  $\langle\left(A_{p}\right)_{\alpha}^{\eta}\rangle=\langle\left(B_{p}\right)_{\alpha}^{\eta}\rangle=0$.
The general formula for  $\Box^{h} R$ -derivative, taking into account  symmetries of   $g_{\mu\nu}$ and its derivatives,  is  \cite{staro}:
\begin{multline}\label{derivordsupsimm}
\frac{\partial \Box^{h}R}{\partial g_{\mu\nu,\eta i_{1}\cdots i_{2h+1}}}=g^{j_{2}j_{3}}\cdots g^{j_{2h}j_{2h+1}}g^{ab}g^{cd}\biggl\{\delta_{a}^{(\mu}\delta_{d}^{\nu)}\delta_{c}^{(\eta}\delta_{b}^{i_{1}}\delta_{j_{2}}^{i_{2}}\cdots\delta_{j_{2h}}^{i_{2h}}\delta_{j_{2h+1}}^{i_{2h+1})}\\
-\delta_{a}^{(\mu}\delta_{b}^{\nu)}\delta_{c}^{(\eta}\delta_{d}^{i_{1}}\delta_{j_{2}}^{i_{2}}\cdots\delta_{j_{2h}}^{i_{2h}}\delta_{j_{2h+1}}^{i_{2h+1})}\biggr\}
\end{multline}
We have to verify the condition  $\langle \left(B_{p}\right)_{\alpha}^{\eta}\rangle=0$. Inserting (\ref{derivordsupsimm}) in the l.h.s. of (\ref{formuladerivatesup}) that, in the weak field limit up to the order  $h^{2}$ becomes

\begin{multline}\label{bassen}
\sum_{h=1}^{p}\sum_{j=0}^{2h}\sum_{m=j+1}^{2h+1}\left(-1\right)^{j}\partial_{i_{0}\cdots i_{j}}\left[\sqrt{-g}a_{h}R\frac{\partial\Box^{h}R}{\partial g_{\mu\nu,\eta i_{1}\cdots i_{m}}}\right]g_{\mu\nu,i_{j+1}\cdots i_{m}\alpha}\\
\stackrel{h^{2}} =\sum_{h=1}^{p}\sum_{j=0}^{2h}\left(-1\right)^{j}\sqrt{-g}^{\left(0\right)}a_{h}\partial_{i_{0}\cdots i_{j}}R^{\left(1\right)}\eta^{j_{2}j_{3}}\cdots \eta^{j_{2h}j_{2h+1}}\eta^{ab}\eta^{cd}\biggl\{\delta_{a}^{(\mu}\delta_{d}^{\nu)}\delta_{c}^{(\eta}\delta_{b}^{i_{1}}\delta_{j_{2}}^{i_{2}}\cdots\delta_{j_{2h}}^{i_{2h}}\delta_{j_{2h+1}}^{i_{2h+1})}\\
-\delta_{a}^{(\mu}\delta_{b}^{\nu)}\delta_{c}^{(\eta}\delta_{d}^{i_{1}}\delta_{j_{2}}^{i_{2}}\cdots\delta_{j_{2h}}^{i_{2h}}\delta_{j_{2h+1}}^{i_{2h+1})}\biggr\}h_{\mu\nu,i_{j+1}\cdots i_{2h+1}\alpha}\\
=\sum_{h=1}^{p}\sum_{j=0}^{2h}\left(-1\right)^{j}a_{h}\partial_{i_{0}\cdots i_{j}}R^{\left(1\right)}Q_{\left(\mu\nu\right)}^{\ \ \ \left(\eta i_{1}\cdots i_{2h+1}\right)}h^{\mu\nu}_{\ \ ,i_{j+1}\cdots i_{2h+1}\alpha}
\end{multline}
with
\begin{equation*}
Q_{\left(\mu\nu\right)}^{\ \ \ \left(\eta i_{1}\cdots i_{2h+1}\right)}=\frac{1}{2!\left(2h+2\right)!}\sum_{ \substack{\mu\nu\in \sigma\left({\mu\nu}\right) \\ \eta i_{1}\cdots i_{2h+1}\in\sigma\left(\eta i_{1}\cdots i_{2h+1}\right)}}Q_{\mu\nu}^{\ \ \ \eta i_{1}\cdots i_{2h+1}}
\end{equation*}
and 
\begin{equation*}
Q_{\left(\mu\nu\right)}^{\ \ \ \left(\eta i_{1}\cdots i_{2h+1}\right)}=\delta_{(\mu}^{(\eta}\delta_{\nu)}^{i_{1}}\eta^{i_{2}i_{3}}\cdots\eta^{i_{2h}i_{2h+1})}-\eta_{(\mu\nu)}\eta^{(\eta i_{1}}\eta^{i_{2}i_{3}}\cdots\eta^{i_{2h}i_{2h+1})}
\end{equation*}
where $\sigma{\left(\mu\nu\right)}$ and $\sigma{\left(\eta i_{1}\cdots i_{2h+1}\right)}$ represent  the set of  index permutations in the brackets. Averaging  (\ref{bassen}) by fixing  $\mathbf{k}$ over a suitable spacetime region adopting a harmonic gauge, we get 
\begin{multline}\label{media2}
\langle\sum_{h=1}^{p}\sum_{j=0}^{2h}\left(-1\right)^{j}a_{h}\partial_{i_{0}\cdots i_{j}}R^{\left(1\right)}Q_{\left(\mu\nu\right)}^{\ \ \ \left(\eta i_{1}\cdots i_{2h+1}\right)}h^{\mu\nu}_{\ \ ,i_{j+1}\cdots i_{2h+1}\alpha}\rangle\\
=\sum_{h=1}^{p}\sum_{j=0}^{2h}\frac{1}{2!\left(2h+2\right)!}\left(-1\right)^{j}a_{h}\sum_{ \substack{\mu\nu\in \sigma\left({\mu\nu}\right) \\ \eta i_{1}\cdots i_{2h+1}\in\sigma\left(\eta i_{1}\cdots i_{2h+1}\right)}}\langle\partial_{i_{0}\cdots i_{j}}R^{\left(1\right)}Q_{\mu\nu}^{\ \ \ \eta i_{1}\cdots i_{2h+1}}h^{\mu\nu}_{\ \ ,i_{j+1}\cdots i_{2h+1}\alpha}\rangle
\end{multline}
The average of  (\ref{media2}) is  independent of  index  permutations in the lower  and  upper cases of  $Q_{\mu\nu}^{\ \ \ \eta i_{1}\cdots i_{2h+1}}$, that is 
\begin{equation}\label{media3}
\langle\partial_{i_{0}\cdots i_{j}}\left(-\frac{1}{2}\Box h\right)Q_{\mu\nu}^{\ \ \ \eta i_{1}\cdots i_{2h+1}}h^{\mu\nu}_{\ \ ,i_{j+1}\cdots i_{2h+1}\alpha}\rangle=\frac{1}{2}\sum_{m=2}^{p+2}\left(-1\right)^{j+h}\left(k_{m}^{2}\right)^{h+1}\left(k_{m}\right)^{\eta}\left(k_{m}\right)_{\alpha}\vert A_{m}\vert^{2}
\end{equation}
By substituting  (\ref{media3}) in (\ref{media2}), we get 
\begin{multline}\label{media4}
\langle\sum_{h=1}^{p}\sum_{j=0}^{2h}\sum_{m=j+1}^{2h+1}\left(-1\right)^{j}\partial_{i_{0}\cdots i_{j}}\left[\sqrt{-g}a_{h}R\frac{\partial\Box^{h}R}{\partial g_{\mu\nu,\eta i_{1}\cdots i_{m}}}\right]g_{\mu\nu,i_{j+1}\cdots i_{m}\alpha}\rangle\\
\stackrel{h^{2}}=\sum_{h=1}^{p}\sum_{j=0}^{2h}\left(-1\right)^{j}a_{h}\sum_{m=2}^{p+2}\left(-1\right)^{j+h}\left(k_{m}^{2}\right)^{h+1}\left(k_{m}\right)^{\eta}\left(k_{m}\right)_{\alpha}\vert A_{m}\vert^{2}\\
=\sum_{h=1}^{p}\sum_{m=2}^{p+2}\left(h+\frac{1}{2}\right)a_{h}\left(-1\right)^{h}\left(k_{m}^{2}\right)^{h+1}\left(k_{m}\right)^{\eta}\left(k_{m}\right)_{\alpha}\vert A_{m}\vert^{2}
\end{multline}
Averaging the right term in  (\ref{formuladerivatesup}), we have
\begin{multline}\label{media5}
\langle\frac{1}{4}\sum_{h=1}^{p}a_{h}\Box h \Box^{h} h^{,\eta}_{\ \ \alpha}+\frac{1}{2}\sum_{h=0}^{1}\sum_{j=h}^{p-1+h}\sum_{m=j+1-h}^{p}\left(-1\right)^{h}a_{m}\Box^{m-j}\left(h^{\eta\lambda}-\eta^{\eta\lambda}h\right)_{,i_{h}\alpha}\Box^{j+1-h}h_{,\lambda}^{\ \ i_{h}}\rangle\\
=\sum_{h=1}^{p}\sum_{m=2}^{p+2}\left(h+\frac{1}{2}\right)a_{h}\left(-1\right)^{h}\left(k_{m}^{2}\right)^{h+1}\left(k_{m}\right)^{\eta}\left(k_{m}\right)_{\alpha}\vert A_{m}\vert^{2}
\end{multline}
Finally, by averaging in the weak field limit Eq. (\ref{formuladerivatesup}) and from  (\ref{media4}) and  (\ref{media5}), we obtain:
\begin{equation}
\langle \left(B_{p}\right)_{\alpha}^{\eta}\rangle=0
\end{equation}
A similar argument gives $\langle \left(A_{p}\right)_{\alpha}^{\eta}\rangle=0$. It is 
\begin{multline}
\langle\sum_{h=1}^{p}\sum_{q=0}^{2h+1}\left(-1\right)^{q}\partial_{i_{0}\cdots i_{q}}\left[\sqrt{-g}a_{h}R\frac{\partial\Box^{h}R}{\partial g_{\mu\nu,\eta i_{1}\cdots i_{q}}}\right]g_{\mu\nu,\alpha}\rangle\\
\stackrel{h^{2}}=\frac{1}{2}\sum_{h=1}^{p}\sum_{m=2}^{p+2}a_{h}\left(-1\right)^{h+1}\left(k_{m}^{2}\right)^{h+1}\left(k_{m}\right)^{\eta}\left(k_{m}\right)_{\alpha}\vert A_{m}\vert^{2}
\end{multline}
\begin{multline}
\langle\frac{1}{2}\sum_{h=1}^{p}a_{h}\Box^{h+1}h_{,\lambda}\left(h^{\eta\lambda}-\eta^{\eta\lambda}h\right)_{,\alpha}\rangle\stackrel{h^{2}}=\frac{1}{2}\sum_{h=1}^{p}\sum_{m=2}^{p+2}a_{h}\left(-1\right)^{h+1}\left(k_{m}^{2}\right)^{h+1}\left(k_{m}\right)^{\eta}\left(k_{m}\right)_{\alpha}\vert A_{m}\vert^{2}
\end{multline}
and then averaging  Eq. (\ref{formuladermax}) on the l.h.s.  and r.h.s., in the weak field limit, we have
\begin{equation}
\langle \left(A_{p}\right)_{\alpha}^{\eta}\rangle=0
\end{equation}
that completes our demonstration.

\section{The polarizations of gravitational waves }\label{AppB}
The six polarizations in the solution \eqref{GW1} can be defined in a suitable matrix base. It is 
\begin{align*}
 \epsilon_{\mu\nu}^{\left(+\right)}&=\frac{1}{\sqrt{2}}
\begin{pmatrix}
0 & 0 & 0 & 0\\0 & 1 & 0 & 0\\0 & 0 & -1 & 0\\0 & 0 & 0 & 0
\end{pmatrix}&
 \epsilon_{\mu\nu}^{\left(\times\right)}&=\frac{1}{\sqrt{2}}
\begin{pmatrix}
0 & 0 & 0 & 0\\0 & 0 & 1 & 0\\0 & 1 & 0 & 0\\0 & 0 & 0 & 0
\end{pmatrix}\\
\epsilon_{\mu\nu}^{\left(\text{TT}\right)}&=\qquad
\begin{pmatrix}
1 & 0 & 0 & 0\\0 & 0 & 0 & 0\\0 & 0 & 0 & 0\\0 & 0 & 0 & 0
\end{pmatrix}&
\epsilon_{\mu\nu}^{\left(\text{TS}\right)}&=\frac{1}{\sqrt{2}}
\begin{pmatrix}
0 & 0 & 0 & 1\\0 & 0 & 0 & 0\\0 & 0 & 0 & 0\\1 & 0 & 0 & 0
\end{pmatrix}\\
\epsilon_{\mu\nu}^{\left(1\right)}&=\frac{1}{\sqrt{2}}
\begin{pmatrix}
0 & 0 & 0 & 0\\0 & 1 & 0 & 0\\0 & 0 & 1 & 0\\0 & 0 & 0 & 0
\end{pmatrix}&
\epsilon_{\mu\nu}^{\left(L\right)}&=\qquad
\begin{pmatrix}
0 & 0 & 0 & 0\\0 & 0 & 0 & 0\\0 & 0 & 0 & 0\\0 & 0 & 0 & 1
\end{pmatrix}
\end{align*}
The $+$ and $\times$ are the two standard of General Relativity. The other are related to the position of non-null terms with respect to the trace (T).
See also \cite{arturo1} for another derivation in fourth order gravity.

\end{document}